\documentclass[10pt,aps,prb,twocolumn,amsmath,amssymb,floatfix,superscriptaddress]{revtex4-2}

\usepackage{graphicx,xcolor}
\usepackage{bm}
\usepackage[
colorlinks,
citecolor=blue,
linkcolor=blue,
urlcolor=blue,plainpages=false,pdfpagelabels
]{hyperref}
\usepackage{txfonts}

\newcommand{\ket}[1]{\left| #1 \right\rangle}
\newcommand{\bra}[1]{\left\langle #1 \right|}
\newcommand{\integral}[3]{\int_{#1}^{#2} #3 \hspace{1pt}}
\newcommand{\inter}{\text{inter}}
\newcommand{\intra}{\text{intra}}

\begin{document}

\title{Kerr effect in tilted nodal loop semimetals}

\author{Johan Ekstr\"om}
\email{johan.ekstrom@uni.lu}
\affiliation{Department of Physics and Materials Science, University of Luxembourg, L-1511 Luxembourg}
\author{Eddwi H. Hasdeo}
\affiliation{Department of Physics and Materials Science, University of Luxembourg, L-1511 Luxembourg}
\affiliation{Research Center for Physics, Indonesian Institute of Sciences, South Tangerang, Indonesia}

\author{M. Bel\'{e}n Farias}
\affiliation{Department of Physics and Materials Science, University of Luxembourg, L-1511 Luxembourg}

\author{Thomas L. Schmidt}
\email{thomas.schmidt@uni.lu}
\affiliation{Department of Physics and Materials Science, University of Luxembourg, L-1511 Luxembourg}

\date{\today}

\begin{abstract}
	We investigate the optical activity of tilted nodal loop semimetals. We calculate the full conductivity matrix for a band structure containing a nodal loop with possible tilt in the $x-y$ plane, which allows us to study the Kerr rotation and ellipticity both for a thin film and a bulk material.
	We find signatures in the Kerr signal that give direct information about the tilt velocity and direction, the radius of the nodal loop and the internal chemical potential of the system. These findings should serve as a guide to understanding optical measurements of nodal loop semimetals and as an additional tool to characterize them.
\end{abstract}

\maketitle

\section{Introduction}

Topological phases in materials have attracted substantial interest over the last two decades, with the discovery of topological insulators \cite{Kane05} at the starting point. As the years have passed, this research field has advanced significantly and we now have a large variety of topological phases at our disposal. Apart from topological insulators, this family of materials now includes for instance Weyl and Dirac semimetals, as well as nodal line, loop and chain semimetals \cite{Armitage18,Burkov11,Bzdusek16}. Similarly to Dirac and Weyl semimetals, a nodal loop semimetal displays band crossings in the energy spectrum. However, in contrast to Dirac and Weyl semimetals where crossings occur at a discrete set of points, the crossing points in this case form a continuous loop in the energy-momentum spectrum. Such a nodal loop is depicted in Fig.~\ref{Fig:NodalSpec} (top).

There is by now a large range of materials that have been suggested to host nodal loops. Experimentally, there is evidence that $\text{ZrSiS}$ \cite{Schilling17,Topp17,Chen17}, $\text{PbTaSe}_{2}$ \cite{Bian16}, $ \text{NbAs}_{2}, $ \cite{Shao18}  and  $ \text{YbMnSb}_{2} $ \cite{Qui19} host nodal loops in their spectra. On the theoretical side, $ \text{Cu}_{3}\text{PdN} $ \cite{Yu15}, $ \text{CaAgAs} $ \cite{wang16} and  $ \text{CaAgP} $ \cite{wang16,xu18}, among others, have been predicted to be nodal loop semimetals. However, further experimental studies are required for their validation. Furthermore, experimental and theoretical work has been performed on $ \text{ZrSiSe} $ samples and it has been shown that nodal loop semimetals may serve as a platform for investigating strongly correlated phases in Dirac materials \cite{Shao20}.

An important experimental tool for studying materials and their properties is the magneto-optical Kerr effect (MOKE). In such experiments, a linearly polarized light beam is directed towards a material surface and one measures the reflected light. Depending on the properties of the material, the reflected light may pick up an orientation-dependent phase difference which can then result in an elliptical polarization of the reflected light. The quantity representing this change in polarization is the Kerr rotation.

In recent years, theoretical works have reported large Kerr rotations for a variety of topological materials and have demonstrated that it is possible to obtain information about their properties from the Kerr rotations. In an early study, Tse \textit{et al.} reported a giant Kerr angle in thin-film topological insulators \cite{Tse10}. This prediction has since been extended to other topological phases \cite{Pratama20}. Kerr rotations have been studied in Refs.~\cite{Kargarian2015, Sonowal19} for different parameter regimes and it was found that Weyl semimetals show signatures of large Kerr rotations as well. Furthermore, a very recent paper by Parent \textit{et al.} shows how the Kerr effect in Weyl semimetals is affected by magnetic fields and demonstrates how a valley polarization and the chiral anomaly can be observed in the Kerr angle \cite{Parent20}.

Some aspects of the optical responses of nodal loop semimetals have already been reported in Refs.~\cite{Barati17, Ahn17, Carbotte16, Ruiz19}. However, Kerr and Faraday effects have not been reported for nodal loop semimetals primarily because of a vanishing Hall response when the nodal loop is untilted. In this paper we investigate in particular how semimetals with a \emph{tilted} nodal loop can be characterized by the Kerr effect. For this purpose, we investigate the Kerr signal both of a thin film of such a material as well as of the bulk material. Tilted nodal loop semimetals have been observed in $\text{ZrSiS}$ and $\text{ZrSiSe}$ \cite{Schilling17, Topp17, Chen17, Shao20} whose nodal loop energy form sinusoidal shapes. Here, we focus on linearly tilted nodal loop semimetals that can be observed by breaking time-reversal symmetry (due to an external magnetic field or internal magnetization) \cite{Huang17}.

Once tilt is introduced a Kerr signal is obtained and we show how this can give information about the tilt of the nodal loop in relation to the radius of the nodal loop. We further show how the Kerr signal depends on the chemical potential of the system.

The structure of this article is as follows: in Sec.~\ref{Sec:Model}, we present the theoretical model we use to describe  a semimetal with a tilted nodal loop. In Sec.~\ref{Sec:OCT} we then derive its optical conductivity tensor and show how its different components can be interpreted from a physical perspective. This quantity is of direct importance for determining the Kerr rotation. Thereafter, we present the general theory for obtaining the Kerr rotation in Sec.~\ref{Sec:MOKE}, and discuss our findings for both a thin film and the bulk geometry. We present our conclusions in Sec.~\ref{Sec:Conclusion}. Throughout this paper we set $ \hbar = e = 1 $.

\section{Model} \label{Sec:Model}

\begin{figure}
	\centering
	\includegraphics[width=0.48\textwidth]{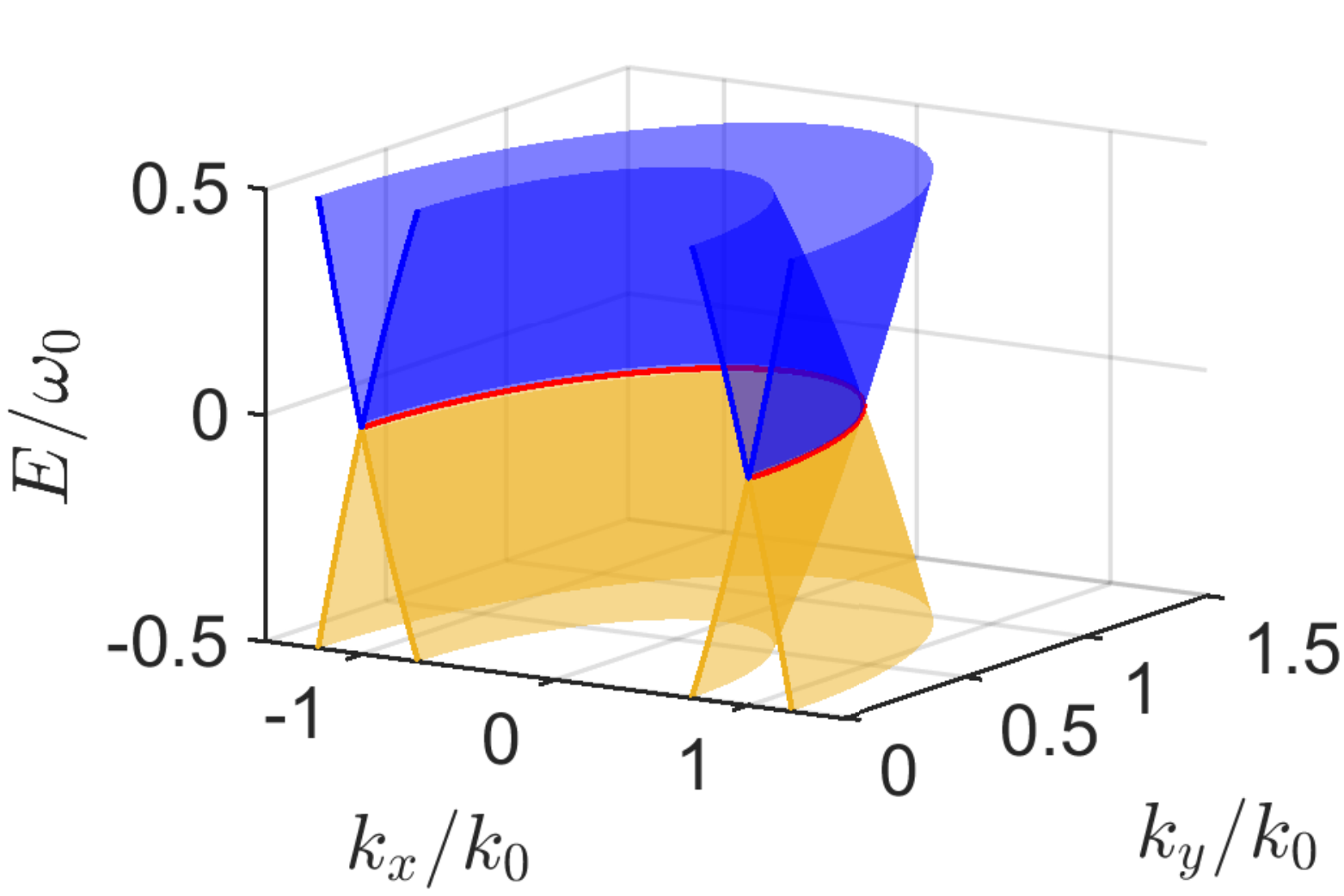}
	\includegraphics[width=0.23\textwidth]{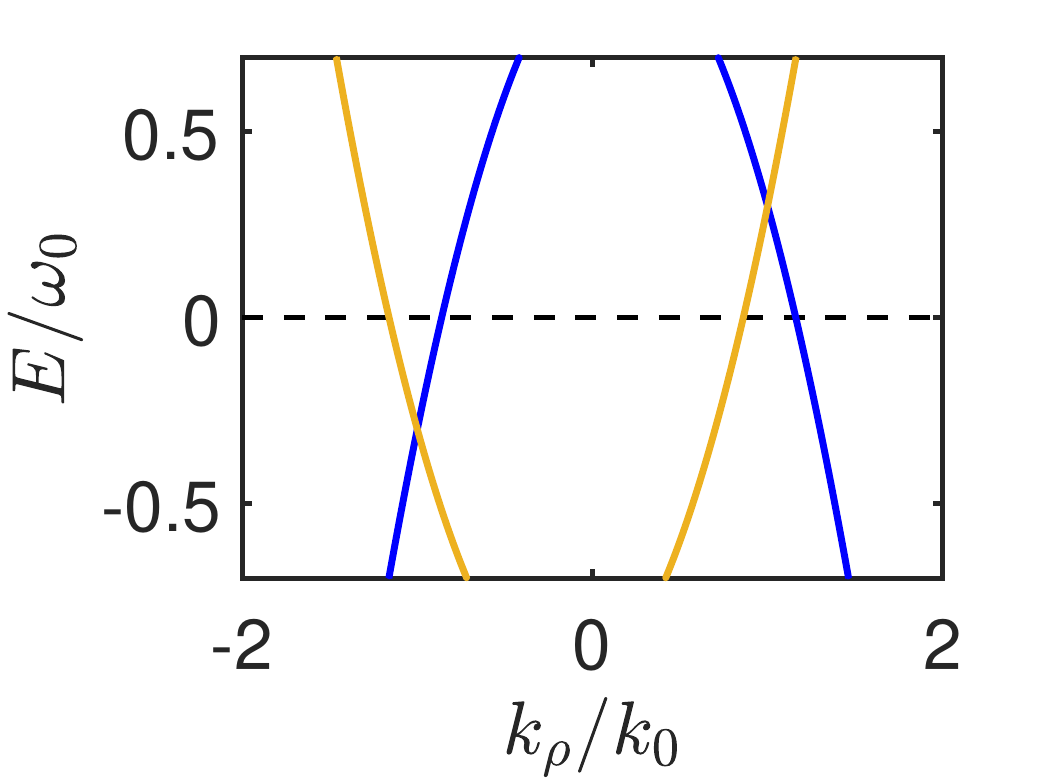}
	\includegraphics[width=0.22\textwidth]{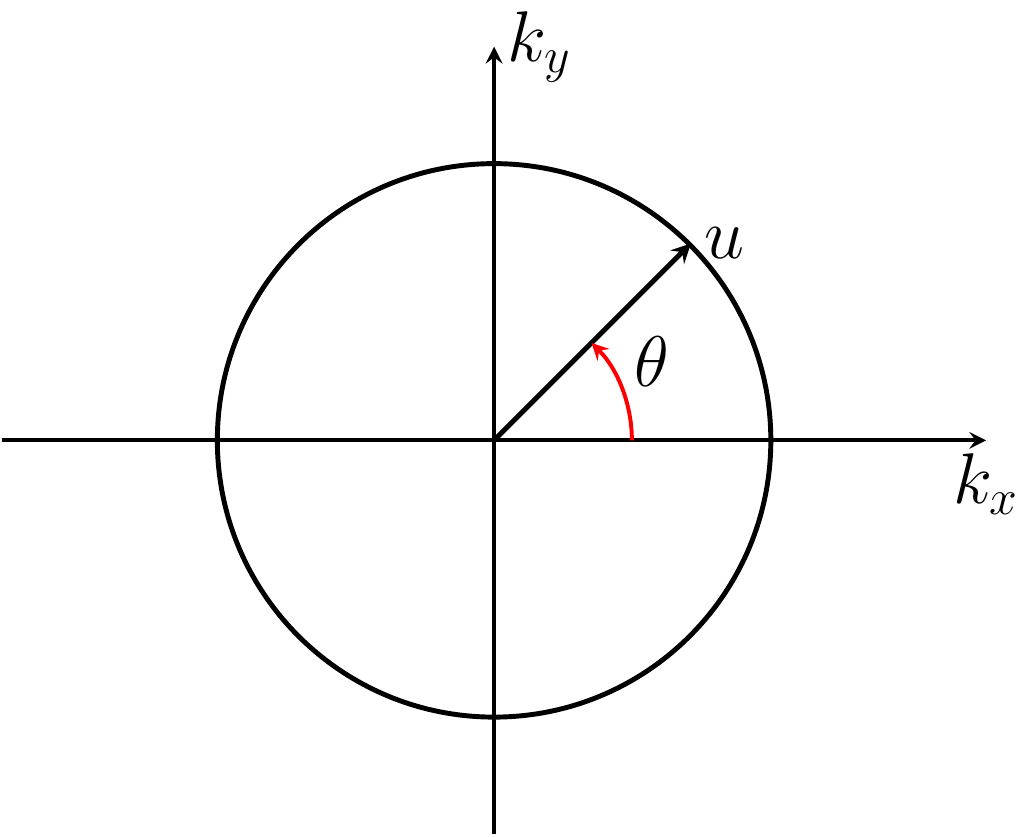}
	\caption{\emph{Top:} A cut through the spectrum of an untilted nodal loop semimetal for $k_z = 0$ and $ \omega_{0} = k_{0}^{2}/\Lambda $. \emph{Bottom left}: Spectrum of a tilted nodal loop semimetal, for $ u =  0.3 k_{0}/\Lambda $ and $ \theta = \pi/4 $. The tilt shifts the nodal loop away from energy $ E=0 $. \emph{Bottom right:} Tilt vector for a tilted nodal loop semimetal.}
	\label{Fig:NodalSpec}
\end{figure}

A nodal line in a band structure naturally emerges at the intersection between two parabolic bands with opposite orientation. At low energies, this model captures the physical properties of a nodal loop semimetal and, taking into account a possible tilt of the nodal loop, the corresponding Hamiltonian for a nodal loop in the $k_x-k_y$ plane is given by \cite{Ruiz19}
\begin{equation}
\hat{H}_{0}(\mathbf{k}) =  \mathbf{u}\cdot \mathbf{k} \tau_{0} + \frac{1}{\Lambda}\left(k_{0}^{2} - k_{\rho}^{2}\right)\tau_{x} + v_{z}k_{z}\tau_{z}.
\label{Eq:Hamiltonian}
\end{equation}
Here, we have defined $k_{\rho}^{2} = k_{x}^{2} + k_{y}^{2}$, $k_0$ is the radius of the nodal loop, $\tau_{x,y,z}$ denotes the vector of Pauli matrices representing two orbital degrees of freedom, $v_z$ is the Fermi velocity in $z$ direction, and $\tau_0$ is the identity matrix. The tilt velocity $\mathbf{u} = (u_{x},u_{y},u_{z})$ causes a tilt of the nodal loop and $\Lambda$ is a mass scale which determines the band curvature and depends on the particular lattice realization used to derive the low-energy Hamiltonian~(\ref{Eq:Hamiltonian}). Its spectrum is given by
\begin{equation}
E_{\mathbf{k},\pm} = \mathbf{u}\cdot\mathbf{k}\pm\sqrt{v_{z}^{2}k_{z}^{2} + \frac{1}{\Lambda^{2}}(k_{0}^{2} - k_{\rho}^{2})^{2} }.
\label{Eq:Spectrum}
\end{equation}
This spectrum is plotted in Fig.~\ref{Fig:NodalSpec} (top) for $\mathbf{u} = 0$. A nodal ring is located in the $E=0$ plane (red line). The radius of this ring is given  by $ k_{0} $. Figure~\ref{Fig:NodalSpec} (bottom left) shows the effect of the tilt. A nonzero tilt shifts the points on the nodal loop away from $E = 0$.

It is convenient to parameterize the tilt velocity vector as
\begin{equation}
\mathbf{u} = (u\cos \theta  ,u\sin \theta  ,u_{z} ),
\end{equation}
where $u = \sqrt{u_{x}^2 + u_{y}^2}$ and $\theta = \arctan(u_{y}/u_{x})$. The angle $ \theta $ represents the tilt direction in the $k_x$-$k_y$ plane and is indicated in Fig.~\ref{Fig:NodalSpec} (bottom right). Moreover, it is straightforward to obtain the following eigenfunctions for the Hamiltonian~\eqref{Eq:Hamiltonian},
\begin{equation}
\Psi_{\pm} = \frac{1}{\sqrt{D_{\pm}}}\begin{pmatrix}
-B \\ v_{z}k_{z} \mp A,
\end{pmatrix}
\end{equation}
where
\begin{align}
    A &= \sqrt{v_{z}^{2}k_{z}^{2} + \frac{1}{\Lambda^{2}}\left(k_{0}^{2} - k_{\rho}^{2}\right)^{2}}, \notag \\
    B &= \frac{1}{\Lambda}\left(k_{0}^{2} - k_{\rho}^{2}\right), \notag \\
    D_{\pm} &= |B|^{2} + \left( v_{z}k_{z} \mp A \right)^{2} .
\end{align}
In the rest of the article, we will consider a nodal loop that is tilted in the $k_x$-$k_y$ plane, so we set $u_{z} = 0$. Furthermore, we will assume the zero-temperature limit when calculating response functions. In the next section, we will present the optical conductivity tensor for the tilted nodal loop system.

\section{Optical conductivity tensor} \label{Sec:OCT}
The coupling between light and matter can be described within the electric dipole approximation, so we consider a Hamiltonian $\hat{H} = \hat{H}_{0} + \mathbf{E}\cdot \mathbf{r}$ where $\mathbf{E}$ is the electric field and $\mathbf{r}$ is the position operator. To obtain the optical conductivity we apply the Kubo formula, which results in
\begin{equation}
\sigma_{ij}(\omega) = \frac{1}{(2\pi)^{3}}\integral{0}{\infty}{dk_{\rho}} k_{\rho}\integral{0}{2\pi}{d\phi}\integral{}{}{dk_{z}} \sigma^{ij}_{\mathbf{k}}(\omega).
\label{Eq:Kubo1}
\end{equation}
where $i,j \in \{x,y,z\}$ and the integral is over all momenta $\mathbf{k}$, written in cylindrical coordinates. The conductivity kernel is defined as
\begin{equation}
\sigma^{ij}_{\mathbf{k}}(\omega) = -i\sum_{s, s^{\prime}}\frac{f(E_{\mathbf{k},s})  - f(E_{\mathbf{k},s^{\prime}})}{E_{\mathbf{k},s} - E_{\mathbf{k},s^{\prime}}}
\frac{j_{\mathbf{k}i}^{ss^{\prime}} j_{\mathbf{k}j}^{s^{\prime}s}}{\omega + E_{\mathbf{k},s} - E_{\mathbf{k},s^{\prime}} + i0^+},
\label{Eq:Kubo2}
\end{equation}
where $f(E)$ denotes the Fermi function and $j_{\mathbf{k}i}^{ss^{\prime}} = \bra{\Psi_{s}}\hat{j}_{\mathbf{k}i}\ket{\Psi_{s^{\prime}}}$ are the optical matrix elements, where $s,s^{\prime} = \pm$ corresponds to the two bands of our model. The conductivity contains contributions both due to transitions between different bands ($ s\neq s^{\prime} $, interband transitions) as well as transitions within a band ($ s=s^{\prime} $, intraband transitions).

The $i$th component of the current operator is defined as $ \hat{j}_{\mathbf{k}i} = \nabla_{\mathbf{k}i}\hat{H} $. For interband transitions we obtain the following matrix elements
\begin{align}
j_{\mathbf{k}x}^{ss^{\prime}} &= \frac{2k_{x} v_{z}k_{z} }{\Lambda A}, \label{Eq:JxInter} \\
j_{\mathbf{k}y}^{ss^{\prime}} &= \frac{2k_{y} v_{z}k_{z} }{\Lambda A}, \label{Eq:JyInter} \\
j_{\mathbf{k}z}^{ss^{\prime}} &= -\frac{v_{z}|B|}{A}, \label{Eq:JzInter}
\end{align}
while for intraband transitions we obtain
\begin{align}
	j_{\mathbf{k}x}^{s} &= u_{x} + \frac{2k_{x} B }{\Lambda A}, \label{Eq:JxIntra} \\
	j_{\mathbf{k}y}^{s} &= u_{y} + \frac{2k_{y} B }{\Lambda A}, \label{Eq:JyIntra} \\
	j_{\mathbf{k}z}^{s} &= \frac{v_{z}^{2}k_{z}}{A}. \label{Eq:JzIntra}
\end{align}
Note that $j_{\mathbf{k}x}^{ss^{\prime}}$ is related to $j_{\mathbf{k}y}^{ss^{\prime}}$ by symmetry in the $x$-$y$ plane. Hence, a symmetry is expected in quantities involving the $x$ and $y$ components.
The real and imaginary parts of Eq.~\eqref{Eq:Kubo2} are obtained by using the identity $\lim_{\gamma \rightarrow 0^+} (x + i\gamma)^{-1} = \mathcal{P}\frac{1}{x} - i\pi\delta(x)$, where $\mathcal{P}$ denotes the Cauchy principal value and $\delta(x)$ is the Dirac delta function. We use this identity to obtain $\text{Re } \sigma_{ij}(\omega)$. From the latter, the imaginary part of the interband contribution is calculated by using the Kramers-Kronig relation,
\begin{equation}
\text{Im } \sigma_{ij}(\omega) = -\frac{1}{\pi}\mathcal{P}\integral{-\infty}{\infty}{d\omega^{\prime}}\frac{\text{Re }\sigma_{ij}(\omega^{\prime})}{{\omega^{\prime}} - \omega}.
\label{Eq:KramersKronig}
\end{equation}
As we will explain below, the contribution arising from the intraband transitions can be directly calculated without applying the Kramers-Kronig relation.

The above equations yield the principal ingredients for calculating the full conductivity tensor. In the $k_\rho$ integral of Eq.~\eqref{Eq:Kubo1}, we make the change of variables $k_{\rho} \rightarrow k_{0}\xi$. Moreover, we introduce the following dimensionless quantities,
\begin{align}
 \tilde{\omega} = \frac{\omega}{\omega_{0}}, \quad \tilde{\mu} = \frac{\mu}{\omega_{0}}  \quad \tilde{u} = \frac{\Lambda u}{k_{0}}, \quad \Gamma = \frac{k_{0}}{v_{z}\Lambda},
\end{align}
where $\mu$ is the chemical potential and $ \omega_{0} = k_{0}^{2}/\Lambda $.

\subsection{Qualitative physical picture}

An analytical calculation of the conductivity tensor~(\ref{Eq:Kubo1}) is in general not possible. However, some essential features of the conductivity tensor and the resulting Kerr response in nodal-line semimetals can be understood based on the symmetry of the conductivity tensor and the form of the current operator $\hat{j}_{\mathbf{k}i}$ alone.

Firstly, one finds that the Hall conductivity $\sigma_{zy}$ vanishes: $E_{\mathbf{k},s}$ is a symmetric function of $k_z$ whereas the product $j^{ss'}_{\mathbf{k}z} j^{s's}_{\mathbf{k}y}$ is odd in $k_z$. By virtue of Eq.~\eqref{Eq:Kubo1} this entails $\sigma_{zy} = 0$. The physical interpretation is that an electric field polarized along the $y$ axis has equal probabilities of exciting electrons carrying currents in the $+z$ and $-z$ directions, so the total current vanishes. By symmetry, the same argument leads to $\sigma_{zx} = \sigma_{xz} = \sigma_{yz} = 0$.

\begin{figure}[t]
	\includegraphics[height=3.95cm]{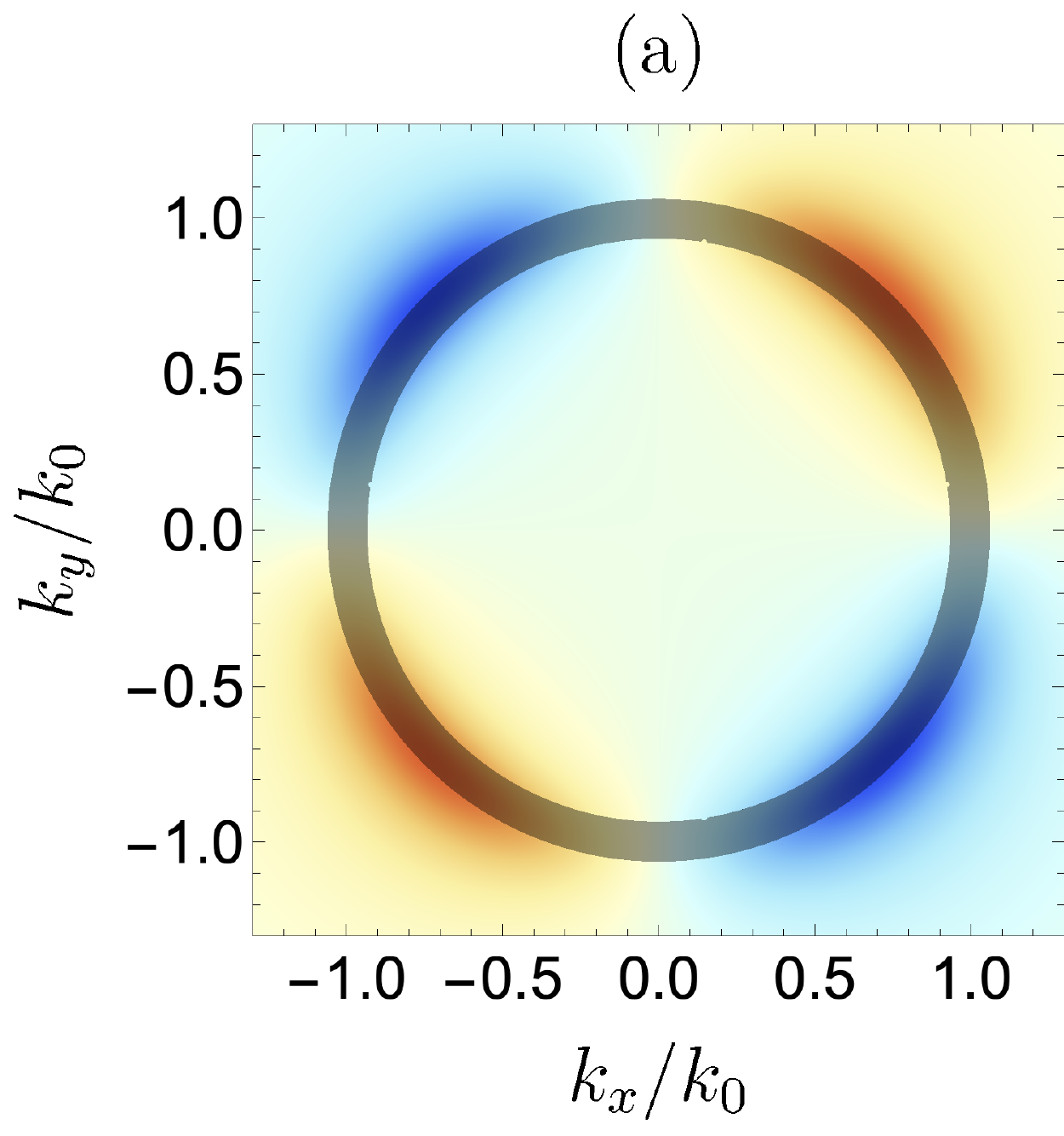}
	\includegraphics[height=3.95cm]{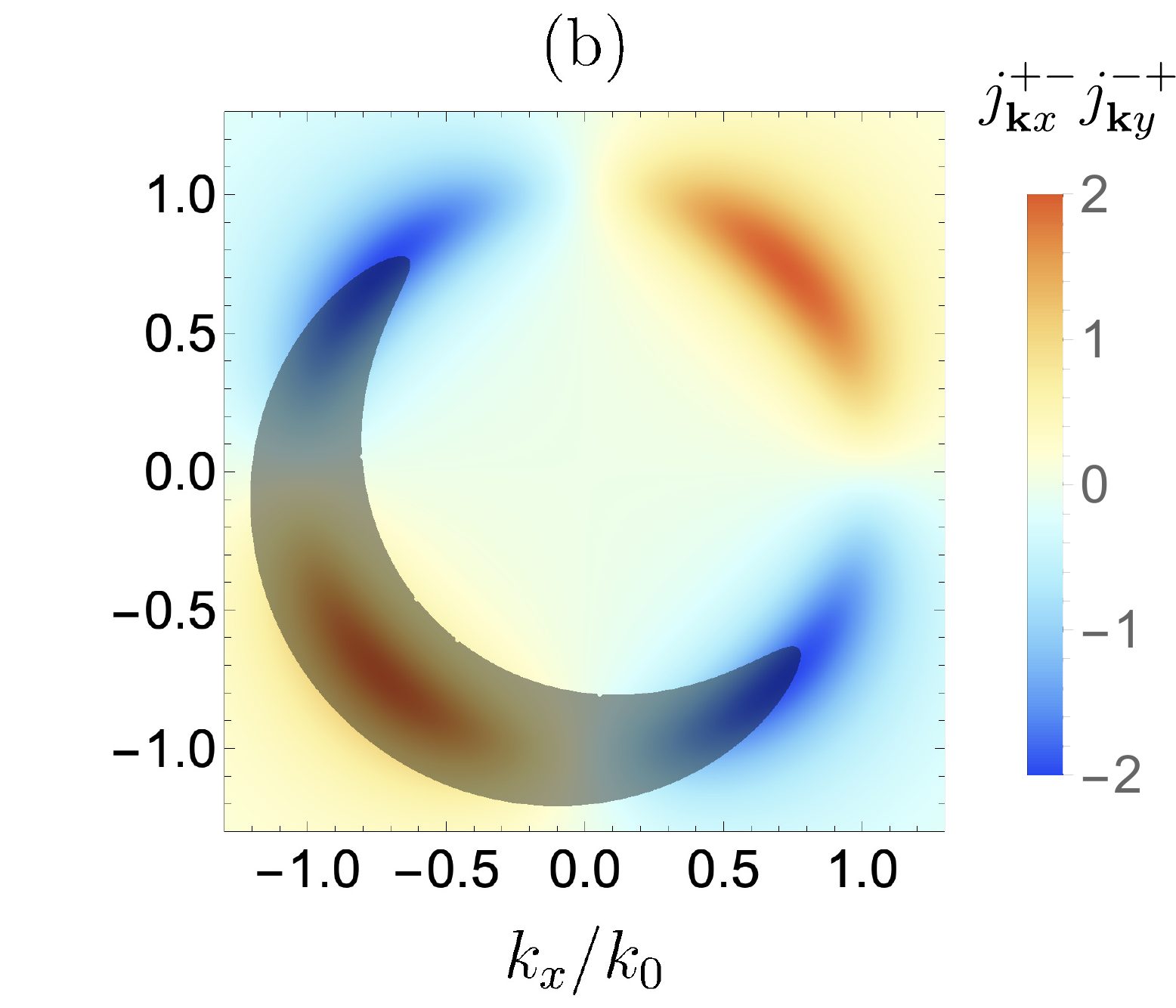}
	\caption{The figure displays the product of $ j_{\mathbf{k}x}^{+-}j_{\mathbf{k}y}^{-+} $, in units of $ k_{0}^{2}/\Lambda^{2} $, for a fixed $ v_{z}k_{z}/\omega_{0}  = 0.4 $. The shaded area displays $ f(E_{\mathbf{k},+}) - f(E_{\mathbf{k},-}) = 0 $ for (a) $ \tilde{u} = 0 $ and for (b) $ \tilde{u} = 0 $ and $\theta = \pi/4$, for a fixed chemical potential $\tilde{\mu} = 0.42$.}
	\label{Fig:XY_current_map}
\end{figure}

The result for $\sigma_{xy}$ is more interesting. In this case, the integrand of Eq.~(\ref{Eq:Kubo1}) becomes an even function of $k_z$. A vanishing conductivity may still arise due to an anti-symmetry of the integrand in Eq.~(\ref{Eq:Kubo1}) in the $k_x$-$k_y$ plane. Indeed, in the absence of tilt ($\mathbf{u} = 0$), the integrand (\ref{Eq:Kubo2}) is anti-symmetric under the change of polar angle $\phi \to \phi + \pi/2$, leading to a vanishing Hall conductivity $\sigma_{xy}$. This anti-symmetry is clearly observed in  Fig.~\ref{Fig:XY_current_map}(a), where we plot the product of $ j_{\mathbf{k}x}^{+-} $ and $ j_{\mathbf{k}y}^{-+} $ as well as $ f(E_{\mathbf{k},+}) - f(E_{\mathbf{k},-}) = 0 $ (visualized by the shaded area) for a fixed $ k_{z} $. The shaded region has a finite thickness even at $\tilde u=0$ because the spectrum is gapped at $k_z\neq 0$. In contrast, if the spectrum is tilted in the $k_x$-$k_y$ plane ($\mathbf{u} \neq 0$), the term containing the Fermi functions in the integrand (\ref{Eq:Kubo2}) will break this polar symmetry and thus make a non-vanishing result for $\sigma_{xy}$ possible (see Fig.~\ref{Fig:XY_current_map}(b)). Therefore, the difference in occupation probabilities of electronic states with momenta $(k_\rho \cos \phi, k_\rho \sin \phi, k_z)$ and $(-k_\rho \cos \phi, -k_\rho \sin \phi, k_z)$ induced by the tilt is at the origin of the nonzero Hall conductivity $\sigma_{xy}$.

The dependence of the Hall conductivities on field polarization and tilt have important consequences for the Kerr signal. If linearly polarized light is incident along the $x$ axis, the electric field will oscillate in the $y$-$z$ plane. Since $\sigma_{yz} = 0$ the main contribution to the current in the material comes from the diagonal elements $\sigma_{yy}$ and $ \sigma_{zz} $. There is a possibility that $ \sigma_{xy} $ contributes to this current but it will be suppressed by $ \sigma_{xx} $. Therefore, the reflected light will be of similar polarization as that of the incident beam. The same argument applies for the case of incident light along the $y$ axis.

In contrast, let us consider the case of light incident along the $z$ axis and, to be specific, let us assume that it is linearly polarized along the $y$ axis. If in addition to the nonzero diagonal term $\sigma_{yy}$, the tilt produces a nonzero $\sigma_{xy}$, the induced current in the material will have components in both $x$ and $y$ directions. The resulting reflected light can therefore be circularly polarized in the $x$-$y$ plane. Therefore, a Kerr effect may be expected for incidence along the $z$ axis.

In the following sections, we will put these qualitative arguments on a solid footing by calculating explicitly the conductivity tensor and the results for the Kerr angle and the Kerr ellipticity.

\subsection{Interband transitions - Real part}
Concerning interband transitions, we find by evaluating the $ k_{z} $ and $ \phi $ integrals of Eq.~\eqref{Eq:Kubo1} that the contributions to the real part of the optical conductivity can be collected in the following two integrals
\begin{align}
\text{Re } \sigma_{ij}^{\inter}(\omega) = \integral{k_{1}}{k_{2}}{d\xi} \Theta(1 - \tilde{\omega}/2) G_{ij}(\xi) \nonumber \\
 + \integral{0}{k_{2}}{d\xi} \Theta(\tilde{\omega}/2 - 1) G_{ij}(\xi),
\end{align}
where $G_{ij}(\xi) = g_{ij}^{+}(\xi) - g_{ij}^{-}(\xi)$ (for $i,j \in \{x,y,z\}$) and $\Theta(x)$ denotes the Heaviside step function. The functions $g_{ij}^\pm(\xi)$ to be integrated are given in the App.~\ref{app:realpart} in Eqs.~(\ref{Eq:fxx})-(\ref{Eq:fzz}). Furthermore, the integration bounds $k_{1} = \sqrt{1 - \tilde{\omega}/2}$ and $k_{2} = \sqrt{1 + \tilde{\omega}/2}$ result from a Dirac-$\delta$ function in the integrand and reflect energy conservation. For a vanishing tilt velocity ($\tilde{u} = 0$), these expressions can be simplified such that analytical results become possible. We have verified that we obtain the same results as Ref.~\cite{Barati17} in this limit.

\begin{figure}
	\includegraphics[width=0.23\textwidth]{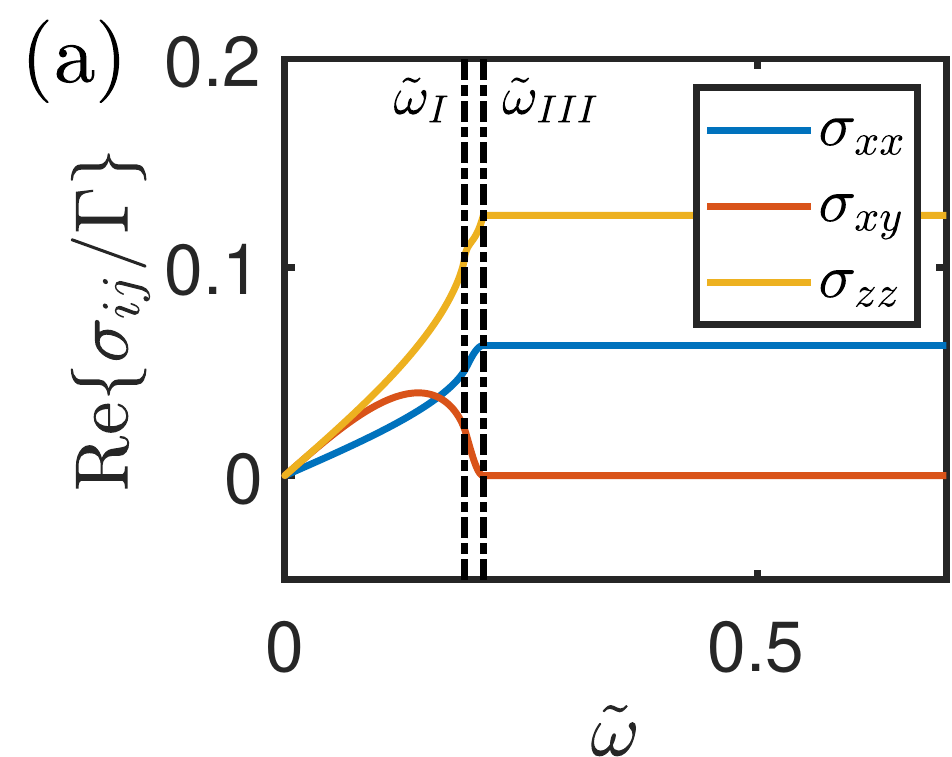}
	\includegraphics[width=0.23\textwidth]{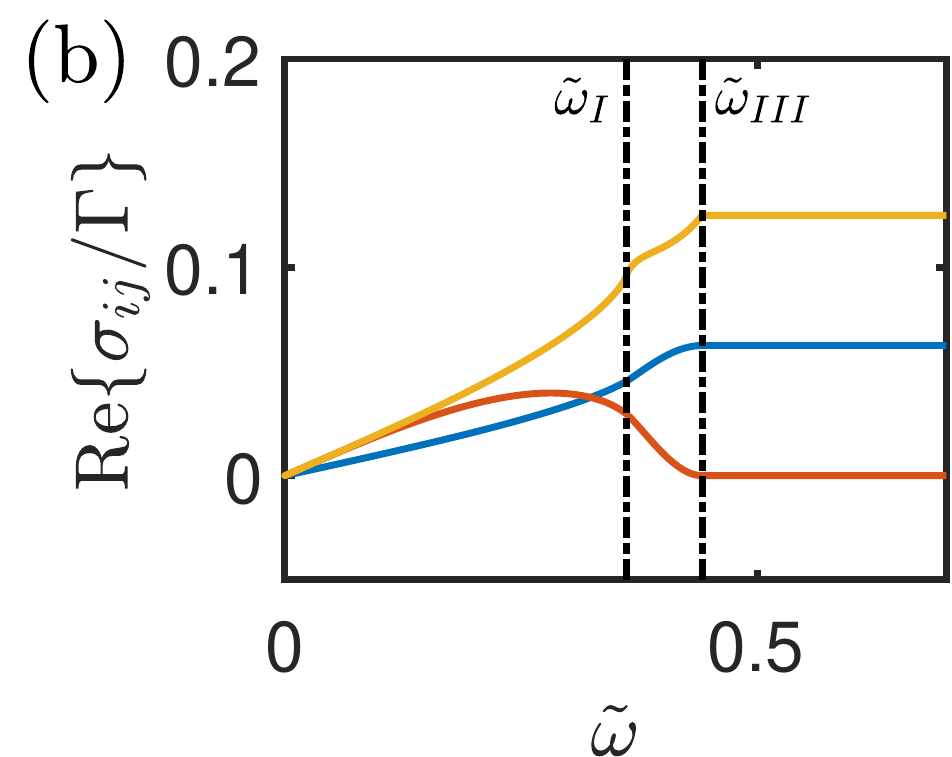}
	\includegraphics[width=0.23\textwidth]{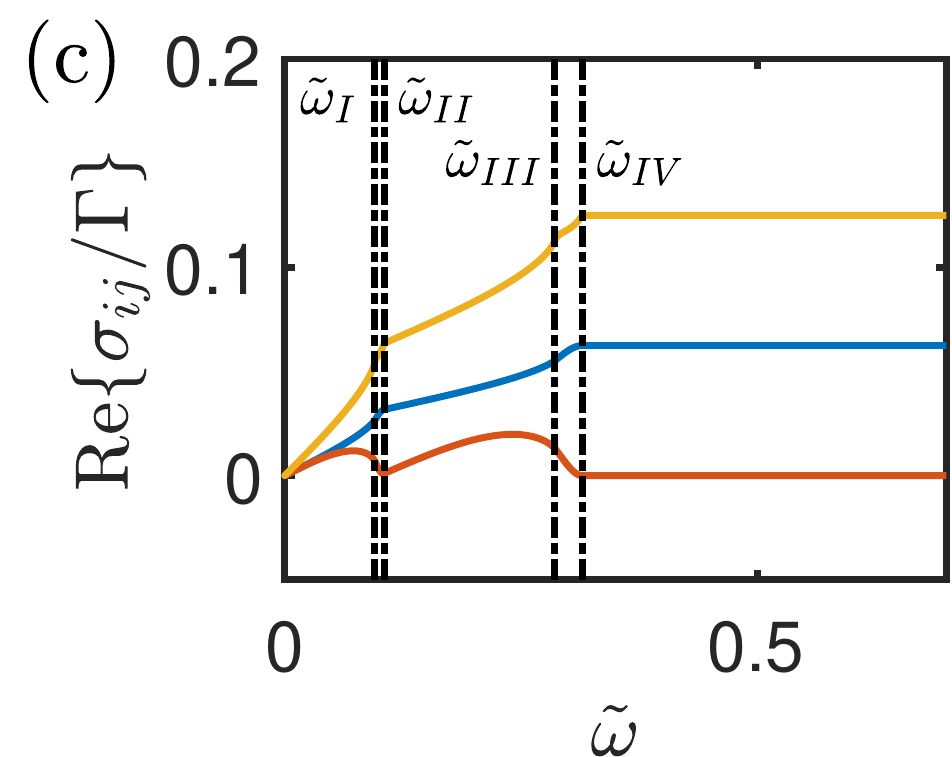}
	\includegraphics[width=0.23\textwidth]{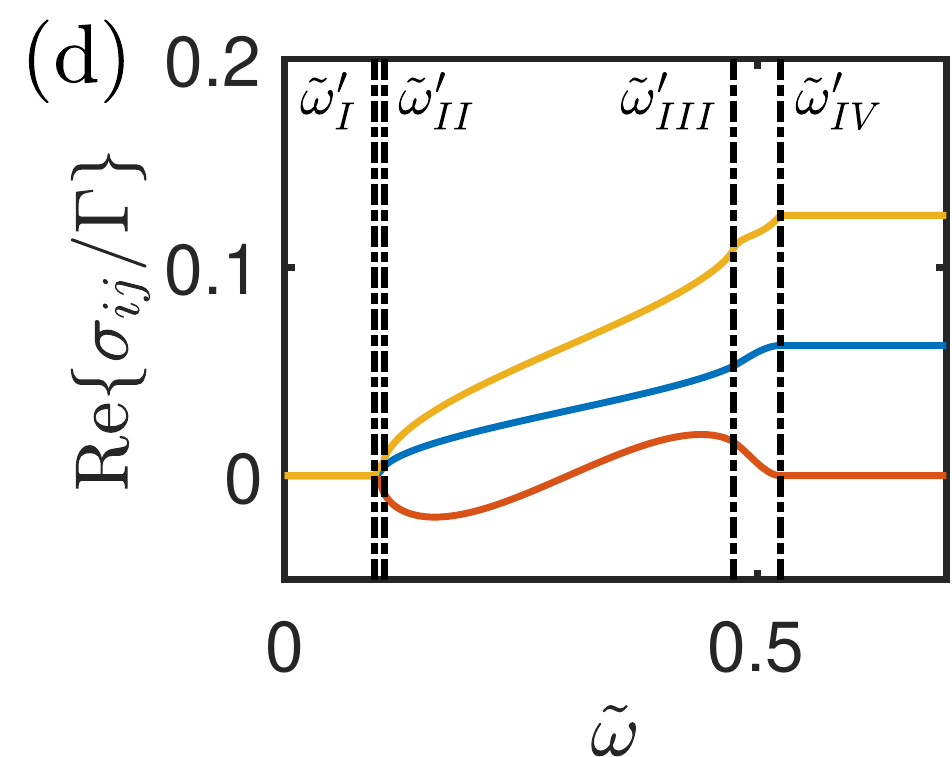}
	\includegraphics[width=0.23\textwidth]{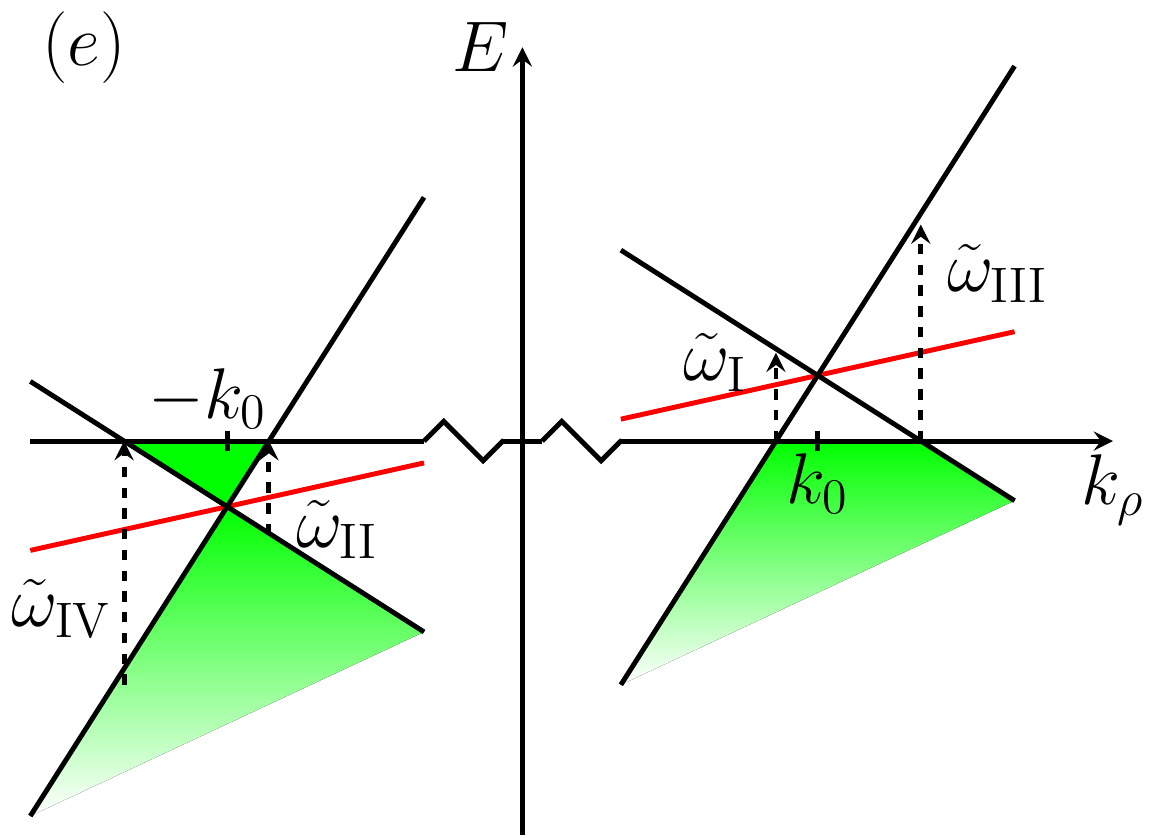}
	\includegraphics[width=0.23\textwidth]{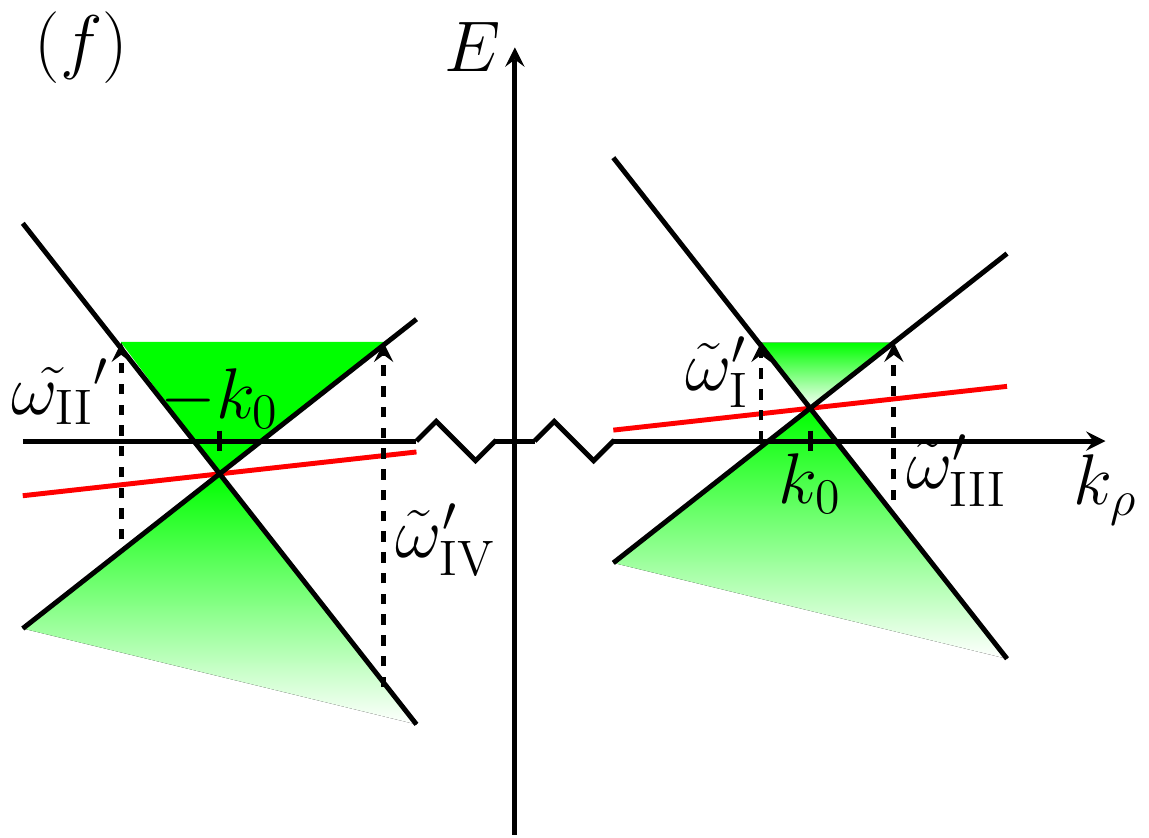}
	\caption{Real part of the optical conductivity for varying $ \tilde{u} $ and $ \tilde{\mu} $. (a) $ \tilde{u} = 0.1 $ and $ \tilde{\mu} = 0 $, (b) $ \tilde{u} = 0.2 $ and $ \tilde{\mu} = 0 $, (c) $ \tilde{u} = 0.1 $ and $ \tilde{\mu} = 0.05 $, (d) $ \tilde{u} = 0.1 $ and $ \tilde{\mu} = 0.15 $. For all plots $ \Gamma = 1 $. The vertical lines show the different transition thresholds as specified in the text and depicted in (e) and (f).  (e) Possible transitions along the nodal loop for $ \tilde{\mu} < \tilde{u} $ and (f) possible transitions for $ \tilde{\mu} > \tilde{u} $. }
	\label{Fig:sR}
\end{figure}

In certain parameter regimes, it is possible to make approximations to obtain analytical expressions. In the limit $\tilde{\omega} \ll 1$, $\tilde{\omega} < \tilde{u}$ and $\tilde{\mu} = 0$ one finds, to linear order in $\tilde{\omega} $, the following results for the real parts of the longitudinal conductivities $\sigma_{ii}$ and the Hall conductivity $\sigma_{xy}$,
\begin{align}
	\text{Re } \sigma_{xx}^{\inter}(\omega)  &\approx \frac{5\Gamma}{24\pi^{2}\tilde{u}}\left (1 - \cos 2\theta\right )\tilde{\omega}, \label{Eq:app_sxx} \\
	\text{Re } \sigma_{yy}^{\inter}(\omega)   &\approx \frac{5\Gamma}{24\pi^{2}\tilde{u}}\left (1 + \cos2\theta \right )\tilde{\omega}, \label{Eq:app_syy} \\
	\text{Re } \sigma_{zz}^{\inter}(\omega)  &\approx \frac{5}{12\pi^{2} \tilde{u}\Gamma}\tilde{\omega}, \label{Eq:app_szz} \\
    \text{Re } \sigma_{xy}^{\inter}(\omega)  &\approx \frac{5\Gamma\sin 2\theta}{12\pi^{2} \tilde{u}} \tilde{\omega}. \label{Eq:app_sxy}
\end{align}
The Hall conductivities along the other transverse directions, $\sigma_{xz}$ and $\sigma_{yz}$, vanish because of our assumption that the tilt is in the $k_x$-$k_y$ plane ($u_{z} = 0$). The next-to-leading order term is proportional to $ \tilde{\omega}^{3} $. This correction becomes important only when $ \theta \approx  n\pi/2$ ($ n\in\mathbb{Z} $) where the linear-order term in $ \sigma_{xx}^{\inter} $ and $ \sigma_{yy}^{\inter} $ can vanish. For $\theta = 0$, our results reproduce the known scaling of the longitudinal conductivities with energy, i.e., $\sigma_{xx}^{\inter} \propto \omega^{3} $ and $ \sigma_{yy}^{\inter} \propto \omega $ \cite{Ahn17}. For general parameters, we have solved the integrals numerically and obtained the results shown in Fig.~\ref{Fig:sR}. The analytical approximations agree with the numerical calculations within the validity range of the approximations.

From the analytical expressions we see that the optical conductivity depends strongly on the tilt angle $\theta$. This applies in particular to $\sigma_{xy}^{\inter}$ which vanishes for $\theta = n \pi/2$ where $n\in \mathbb{Z}$. At these tilt angles, the electric field excites an equal flow of electrons in opposite directions which hence cancel each other. This is also observed for the dc Hall conductivity \cite{Ruiz19}. Furthermore, an untilted nodal loop ($\tilde{u} = 0$) does not give rise to a Hall conductance, as was already shown for the dc Hall conductivity in Ref.~\cite{Ruiz19}. For the ac Hall conductivity this follows from the fact that the integral over the angular coordinate in Eq.~\eqref{Eq:Kubo1} separates and vanishes as $  \int_{0}^{2\pi}d\phi\sin\phi\cos\phi  = 0$, see also the matrix elements defined in Eqs.~\eqref{Eq:JxInter} and \eqref{Eq:JyInter}.

Secondly, we observe that $\sigma_{xy}^{\inter}$ reaches its maximum for $\theta = n\pi/4$, with odd $n$, in which case we also obtain $\sigma_{xx} = \sigma_{yy}$. To be specific, let us discuss the case $\theta = \pi/4$. The numerical results are plotted in Fig.~\ref{Fig:sR}. Note that in all cases we are considering $0<\tilde{u},\tilde{\mu}<1$ without losing generality. 
To investigate the physical processes giving rise to the conductivity for $\tilde{\mu} < \tilde{u}$, one can define the following threshold energies,
\begin{align}
\tilde{\omega}_{\text{I}} &= -2\tilde{\mu} - \tilde{u}^{2} + \tilde{u}\sqrt{\tilde{u}^{2} + 4(1 + \tilde{\mu})}, \notag \\
\tilde{\omega}_{\text{II}} &= 2\tilde{\mu} - \tilde{u}^{2} + \tilde{u}\sqrt{\tilde{u}^{2} + 4(1 - \tilde{\mu})}, \notag \\
\tilde{\omega}_{\text{III}} &= -2\tilde{\mu} + \tilde{u}^{2} + \tilde{u}\sqrt{\tilde{u}^{2} + 4(1 - \tilde{\mu})}, \notag \\
\tilde{\omega}_{\text{IV}} &= 2\tilde{\mu} + \tilde{u}^{2} + \tilde{u}\sqrt{\tilde{u}^{2} + 4(1 + \tilde{\mu})}.
\end{align}
The thresholds are ordered such that $ \tilde{\omega}_{\text{I}} < \tilde{\omega}_{\text{II}} <\tilde{\omega}_{\text{III}} <\tilde{\omega}_{\text{IV}} $. In the case $ \tilde{\mu} = 0 $ we have $ \tilde{\omega}_{\text{I}} = \tilde{\omega}_{\text{II}} $ and $ \tilde{\omega}_{\text{III}} = \tilde{\omega}_{\text{IV}} $. The threshold energies mark the onset of allowed vertical transitions around the nodal loop. As long as $ \tilde{\omega} < \tilde{\omega}_{\text{IV}} $ the system is partially Pauli blocked. As the energy of the incoming photons increases, more vertical transitions become allowed around the nodal loop. Once the energy of an incoming photon is larger than $ \tilde{\omega}_{\text{IV}} $ vertical transitions are allowed everywhere on the nodal loop and the Pauli blockade has been lifted. The absorption processes defining the threshold energies are depicted in the graph in Fig.~\ref{Fig:sR}(e) and are indicated in the numerical plots in Fig.~\ref{Fig:sR}(a-c).

In the case $\tilde{u} < \tilde{\mu}$ we find that the energy thresholds are given by
\begin{align}
\tilde{\omega}_{\text{I}}^{\prime} &= 2\tilde{\mu} + \tilde{u}^{2} - \tilde{u}\sqrt{\tilde{u}^{2} + 4(1 + \tilde{\mu})}, \notag \\
\tilde{\omega}_{\text{II}}^{\prime} &= 2\tilde{\mu} - \tilde{u}^{2} - \tilde{u}\sqrt{\tilde{u}^{2} + 4(1 - \tilde{\mu})}, \notag \\
\tilde{\omega}_{\text{III}}^{\prime} &= 2\tilde{\mu} - \tilde{u}^{2} + \tilde{u}\sqrt{\tilde{u}^{2} + 4(1 - \tilde{\mu})}, \notag \\
\tilde{\omega}_{\text{IV}}^{\prime} &= 2\tilde{\mu} + \tilde{u}^{2} + \tilde{u}\sqrt{\tilde{u}^{2} + 4(1 + \tilde{\mu})}.
\end{align}
The corresponding allowed transitions are depicted in Fig.~\ref{Fig:sR}(f). Going from $\tilde{u} > \tilde{\mu}$ to $\tilde{u} < \tilde{\mu}$ changes the allowed vertical transitions around the nodal loop. As can be seen from Fig.~\ref{Fig:sR}(d), for $ \omega < \tilde{\omega}_{\text{I}}^{\prime} $ the system is Pauli blocked and no vertical transitions are allowed. Once $ \omega > \tilde{\omega}_{\text{I}}^{\prime} $ the Pauli blockade is overcome and transitions are partially allowed around the nodal loop. We further note the impact of $\tilde{u} < \tilde{\mu}$ has on $ \sigma_{xy}^{\inter} $. It now takes both positive and negative values.

Finally, we note that for $ \tilde{\omega} > \tilde{\omega}_{\text{IV}}, \tilde{\omega}_{\text{IV}}^{\prime}$, transitions from the valence to the conduction band are allowed all around the nodal loop and the response of the system then resembles that of an untilted nodal loop. As has been previously observed, the conductivity in this region reaches a constant value \cite{Ahn17,Barati17}.

\begin{figure}[t]
	\includegraphics[width=0.23\textwidth]{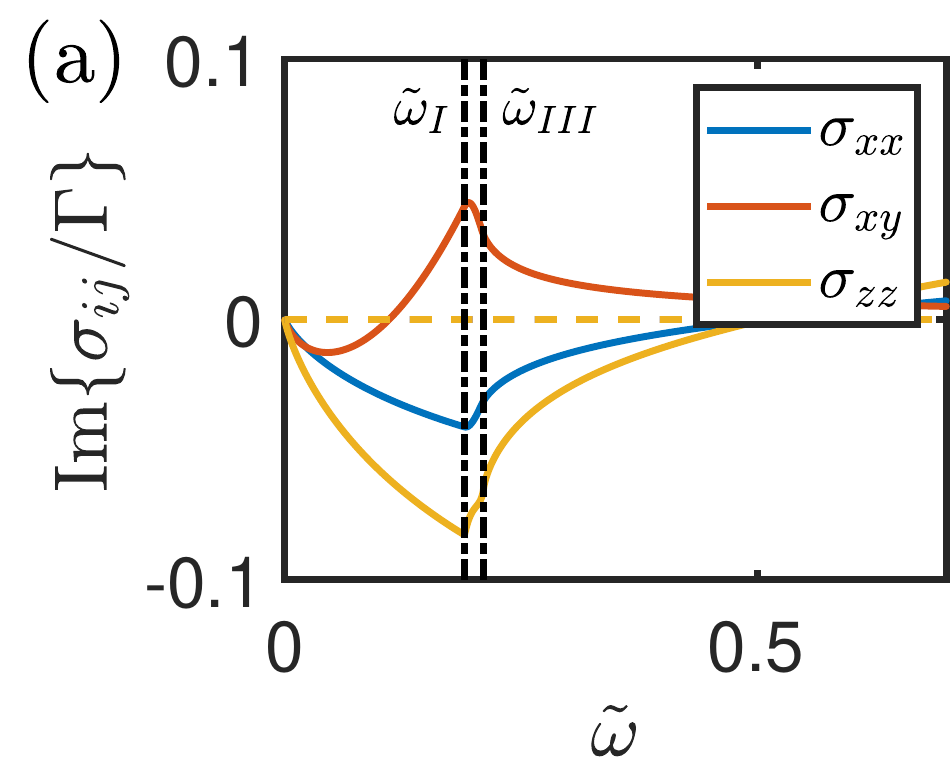}
	\includegraphics[width=0.23\textwidth]{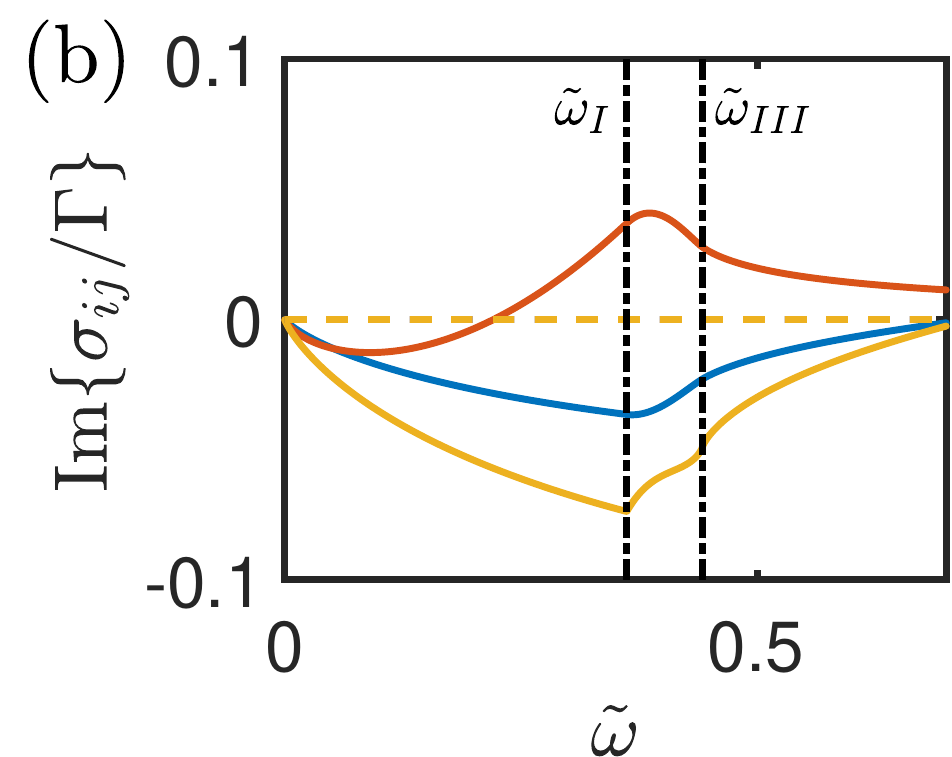}
	\includegraphics[width=0.23\textwidth]{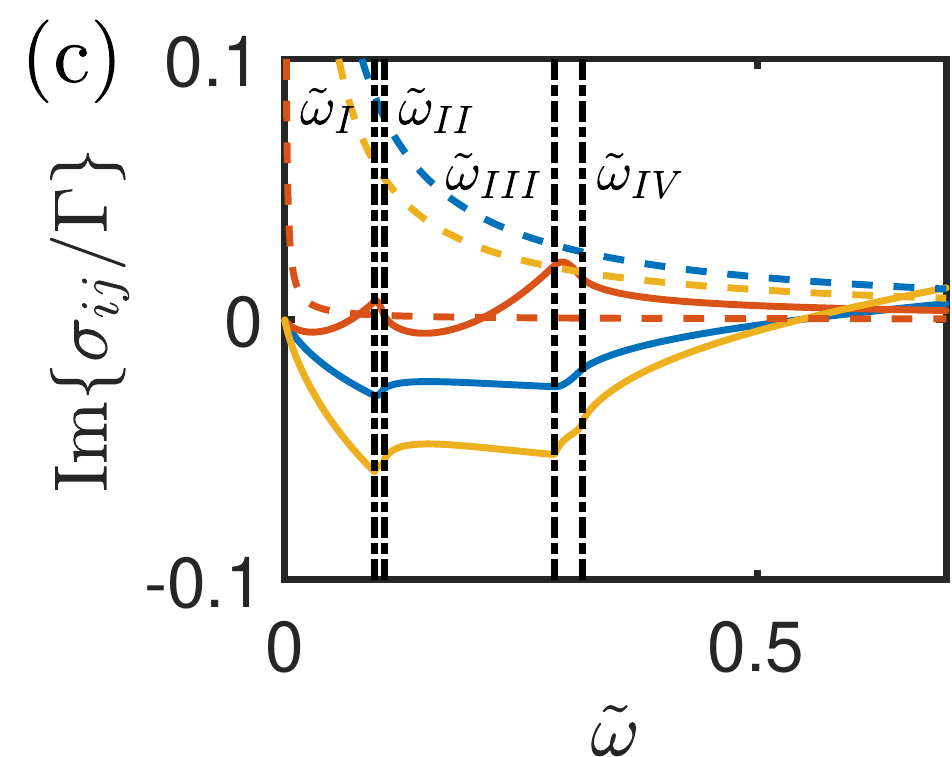}
	\includegraphics[width=0.23\textwidth]{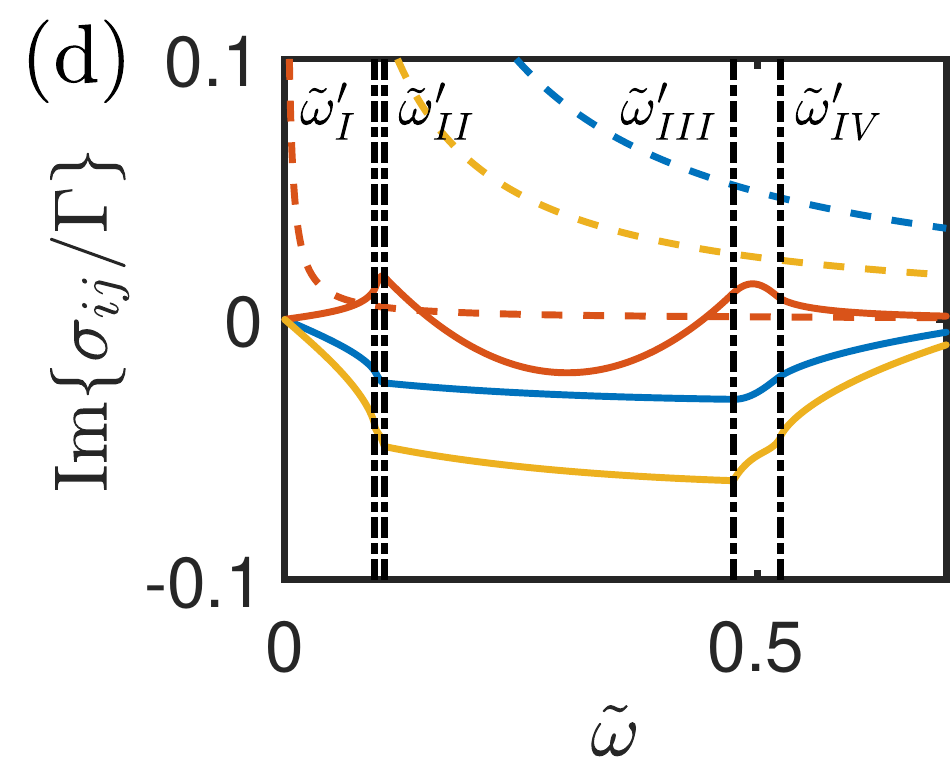}
	\caption{Imaginary part of the optical conductivity for varying $ \tilde{u} $ and $ \tilde{\mu} $. (a) $ \tilde{u} = 0.1 $ and $ \tilde{\mu} = 0 $, (b) $ \tilde{u} = 0.2 $ and $ \tilde{\mu} = 0 $, (c) $ \tilde{u} = 0.1 $ and $ \tilde{\mu} = 0.05 $, (d) $ \tilde{u} = 0.1 $ and $ \tilde{\mu} = 0.15 $. The solid and dashed lines correspond to interband and intraband contributions respectively. The vertical lines show the different transition thresholds as specified in the text and depicted in Fig.~\ref{Fig:sR}(e) and (f). For all plots $ \Gamma = 1 $.}
	\label{Fig:sI}
\end{figure}

\subsection{Interband transitions - Imaginary part}
To obtain the imaginary part of $\sigma_{ij}^{\inter}(\omega)$, we apply the Kramers-Kronig relation~\eqref{Eq:KramersKronig} to the results obtained for the real part of the conductivity tensor. The integration is done numerically. A cutoff $\omega_c$ has to be introduced to regularize a logarithmic divergence in the unbounded integral. The necessity of a cutoff is not surprising because the Kramers-Kronig relation involves an infinite integration range whereas the effective model we consider is only valid for small energies. We have chosen $\omega_{c} = 2k_{0}^{2}/\Lambda$ and have verified that a larger cut-off energy has no pronounced effect on the low-energy conductivity.

The imaginary part of the interband contributions to the conductivity tensor is plotted as solid lines in Fig.~\ref{Fig:sI}. For zero chemical potential the imaginary parts are negative for small $\tilde{\omega}$, and $\sigma_{xx,yy}$ remains so even for larger frequencies. On the contrary, the imaginary part of $\sigma_{xy}$ increases and takes on positive values for increasing $\tilde{\omega}$. Thereafter it decreases towards zero.

\subsection{Intraband transitions}

The intraband transitions are obtained by setting $ s = s^{\prime} $ in Eq.~\eqref{Eq:Kubo2}. In this case the conductivity kernel becomes
\begin{equation}
	\sigma_{\mathbf{k}}^{ij,\intra}(\omega) = -i\sum_{s}\frac{\partial f(E)}{\partial E}\Big|_{E=E_{\mathbf{k},s}}\frac{j_{\mathbf{k}i}^{s}j_{\mathbf{k}j}^{s}}{\omega + i0^{+}}.
\end{equation}
At zero temperature the derivative of the Fermi distribution reduces to a delta function, $ \frac{\partial f(E)}{\partial E} = -\delta(E) $. By using the identity $\lim_{\gamma \rightarrow 0^+} (x + i\gamma)^{-1} = \mathcal{P}\frac{1}{x} - i\pi\delta(x)$ the contribution to the optical conductivity from intraband transitions can be written as
\begin{equation}
	\sigma_{ij}^{\intra}(\omega) = D_{ij}\left( \frac{i}{\omega} +\pi\delta(\omega) \right),
\end{equation}
where the Drude weight, $ D_{ij} $, has the form of
%, after evaluating the $ k_{z} $-integral of Eq.~\eqref{Eq:Kubo1},
\begin{equation}
	D_{ij} = \frac{4}{(2\pi)^{3}}\integral{0}{\infty}{d\xi}\integral{0}{2\pi}{d\phi} \tilde{G}_{ij}(\xi,\phi).
	\label{Eq:IntrabandAmplitude}
\end{equation}
The functions $\tilde{G}_{ij}(\xi,\phi) $ are given in Appendix~\ref{App:Intraband}. The integrals can be calculated analytically for $ \tilde{u} = 0 $ for which we confirm the results of Ref.~\cite{Barati17}. For a finite tilt velocity the integrals must be evaluated numerically.

As the contribution from the intraband transitions to the real part of the optical conductivity is a delta function at $ \tilde{\omega} = 0 $, we neglect this term since we are interested in finite frequencies. The existence of this singularity at $\tilde{\omega} = 0$ can be traced back to the absence of dissipation in our system. However, the imaginary part has a frequency dependence and will be included. We plot it as dashed lines along with the imaginary part of the interband transitions in Fig.~\ref{Fig:sI}. We note that the Drude weight $D_{ij}$ vanishes for $\mu=0$ resulting in no intraband contribution in $\mathrm{Im}\{\sigma_{ij}\}$ as shown in Fig.~\ref{Fig:sI}(a) and (b). As can been seen from the plot, $ D_{xy} $ has a minor impact on $ \text{Im }\sigma_{xy} $ and mainly contributes for small $ \tilde{\omega} $. The intraband contribution has a greater impact on the $ \sigma_{xx/yy/zz} $ as it does not go to zero as fast as $ \sigma_{xy}^{\inter} $.

To summarize this section, we have calculated the full frequency-dependent conductivity tensor for a nodal loop semimetal tilted in the $k_x$-$k_y $ plane. In the following, we will use this quantity for the calculation of the Kerr response.

\section{Magneto optical Kerr effect} \label{Sec:MOKE}

\begin{figure}[t]
\centering
\includegraphics[width=0.49\textwidth]{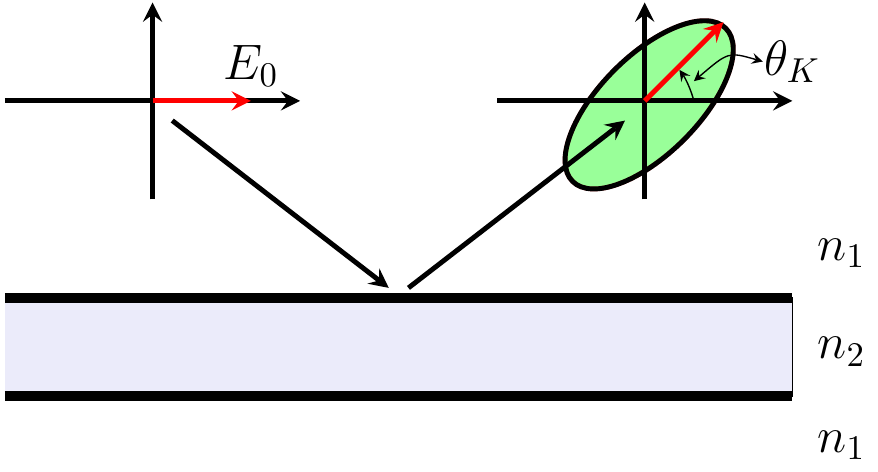}
\caption{Schematic picture illustrating the magneto-optical Kerr effect. A linearly polarized incident light beam is reflected with a elliptic polarization. The latter is characterized by the Kerr angle $\theta_K$ and the Kerr  ellipticity  $\epsilon_K$. }
\label{Fig:KerrRotation}
\end{figure}

The magneto-optical Kerr effect acts as an optical tool for characterizing and understanding different materials. When an incident electromagnetic wave is reflected from the surface of a material, the reflected wave may pick up a polarization-dependent phase, thus corresponding to a change in the polarization angle as well as in an elliptic polarization of the reflected wave. This phenomenon is referred to as the Kerr rotation and is depicted in Fig.~\ref{Fig:KerrRotation}.

The Kerr rotation is governed by the properties of the material, which are represented in Maxwell's equations describing the propagation of light in the vacuum and inside the material. We will consider the Kerr effect when a linearly polarized electromagnetic wave with normal incidence is reflected on the surface of a nodal loop semimetal. Both a free standing thin-film and a semi-infinite bulk material will be considered. The general theory is outlined in the following paragraphs, whereas the specific details that have to be applied for the two different geometries as well as the results will be presented in the following sections.

A linearly polarized incident wave can be represented as an equal superposition of two circularly polarized waves. We will limit ourselves to the case of normal incidence, and we first consider a wave travelling along the $z$ direction. In this case, we can write a circularly polarized wave as $\mathbf{E}^{R,L} = E_{0}^{R,L} \hat{e}_{R,L} e^{i(\mathbf{k}\cdot\mathbf{r} - \omega t)}$, where $\hat{e}_{R,L} = \hat{\mathbf{x}} \mp i\hat{\mathbf{y}}$ and $\mathbf{k} = k_z \hat{\mathbf{z}}$. The basis vectors $\hat{\mathbf{x}}$ and $\hat{\mathbf{y}}$ are in the plane perpendicular to the direction of the wave propagation and the minus and plus signs correspond to right-handed ($ R $) and left-handed ($ L $) circularly polarized light, respectively.

The Kerr angle and ellipticity are now obtained by considering the quotient between the reflection amplitudes of a right- and a left-handed circularly polarized beam with the complex amplitudes $ E_{r}^{R} $ and $ E_{r}^{L} $ \cite{Tse10,Oppeneer99}. This complex quotient has a magnitude and a phase, $E_{r}^{R}/E_{r}^{L} = |E_{r}^{R}|/|E_{r}^{L}|e^{i(\alpha_{R} - \alpha_{L})}$. The phase and the magnitude define, respectively, the Kerr angle and the Kerr ellipticity \cite{Oppeneer99},
\begin{align}
\theta_{K} &= \frac{1}{2}(\alpha_{R} - \alpha_{L}), \label{Eq:KerrAngle} \\
\epsilon_{K} &= \frac{|E_{r}^{R}|}{|E_{r}^{L}|}.
\end{align}
We will now explain how to calculate the reflection amplitudes for a light beam incident on a free-standing thin film and a semi-infinite bulk material.

\subsection{Thin film}

\begin{figure}
	\includegraphics[width=0.23\textwidth]{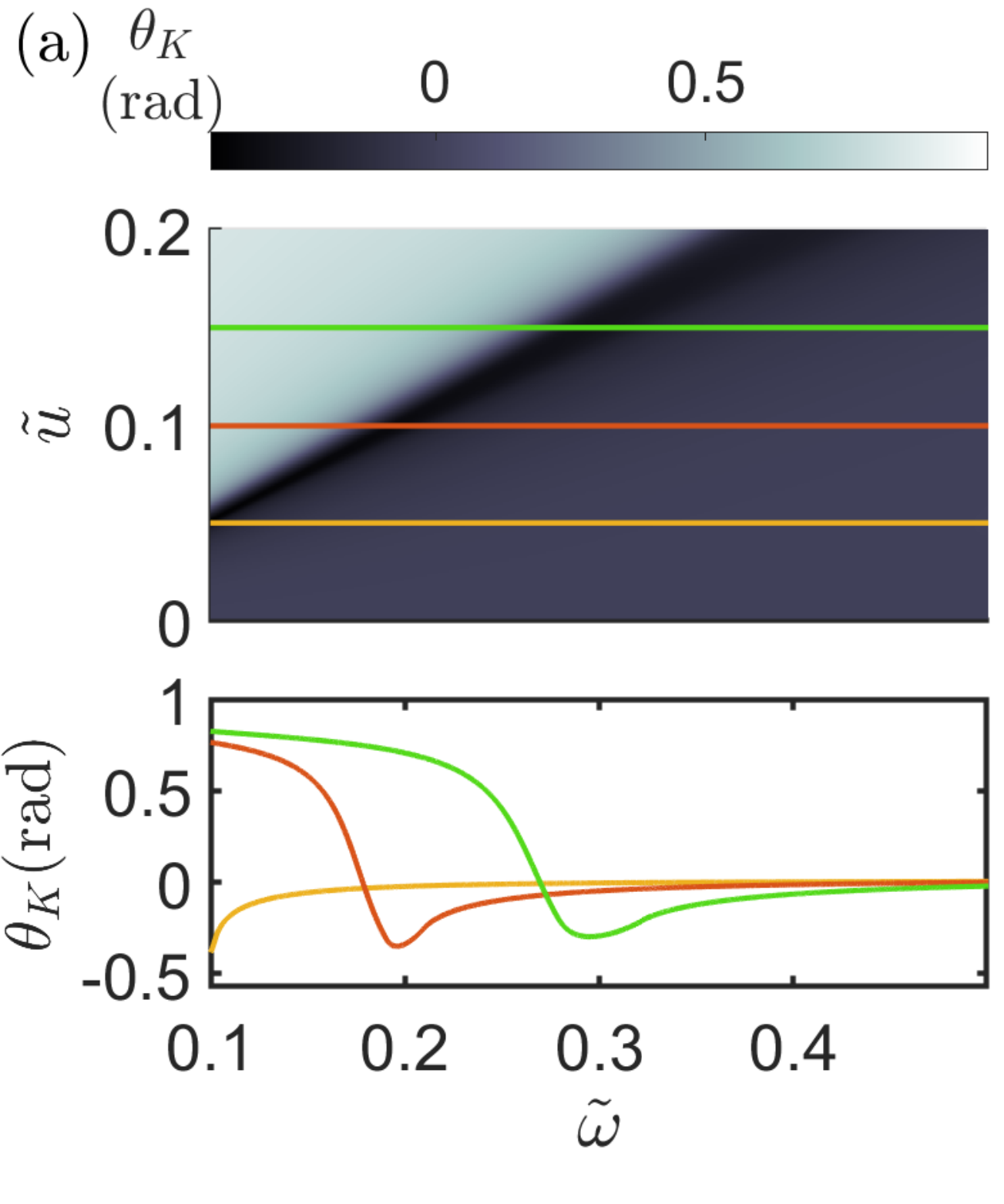}
	\includegraphics[width=0.23\textwidth]{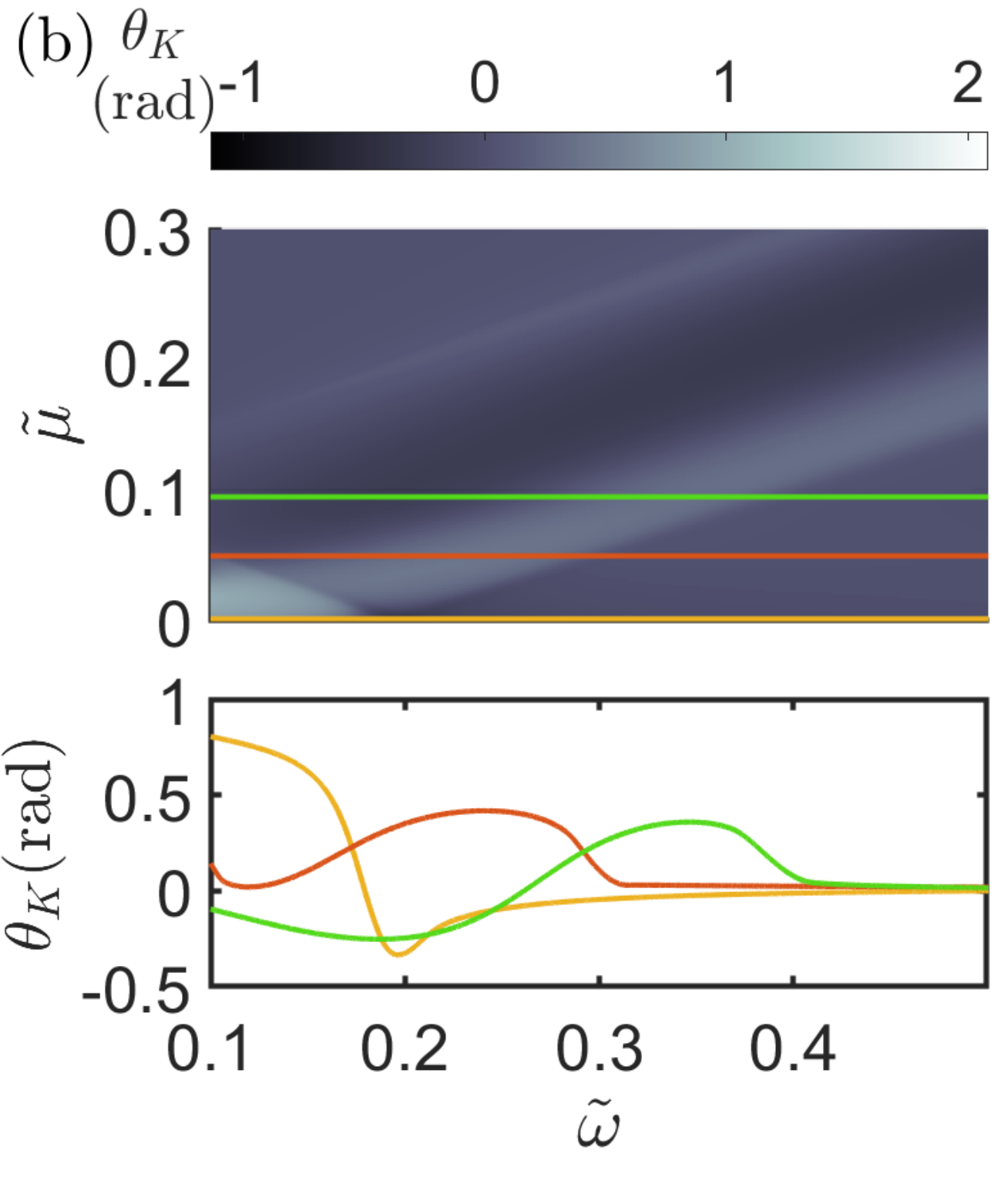}
	\includegraphics[width=0.23\textwidth]{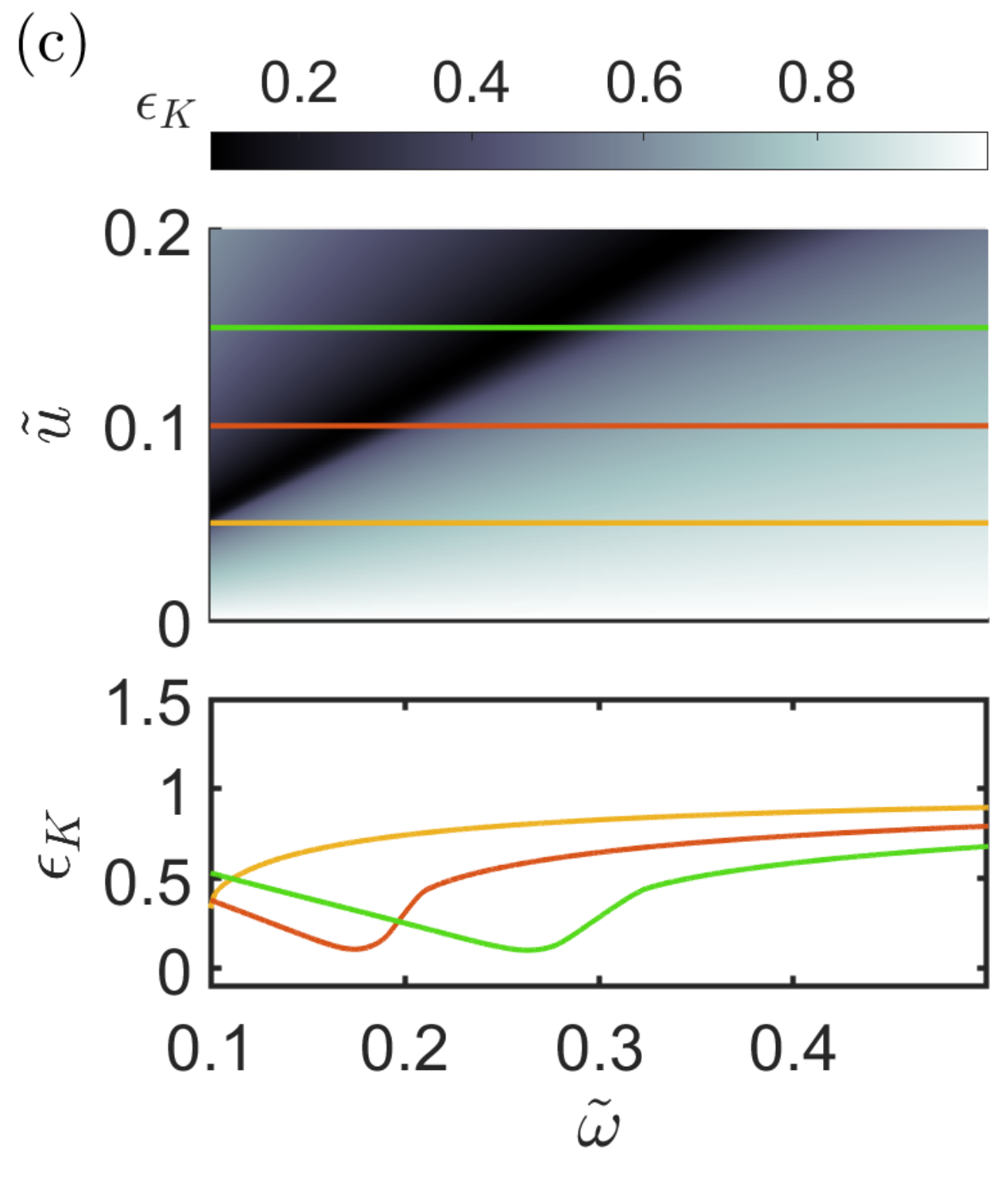}
	\includegraphics[width=0.23\textwidth]{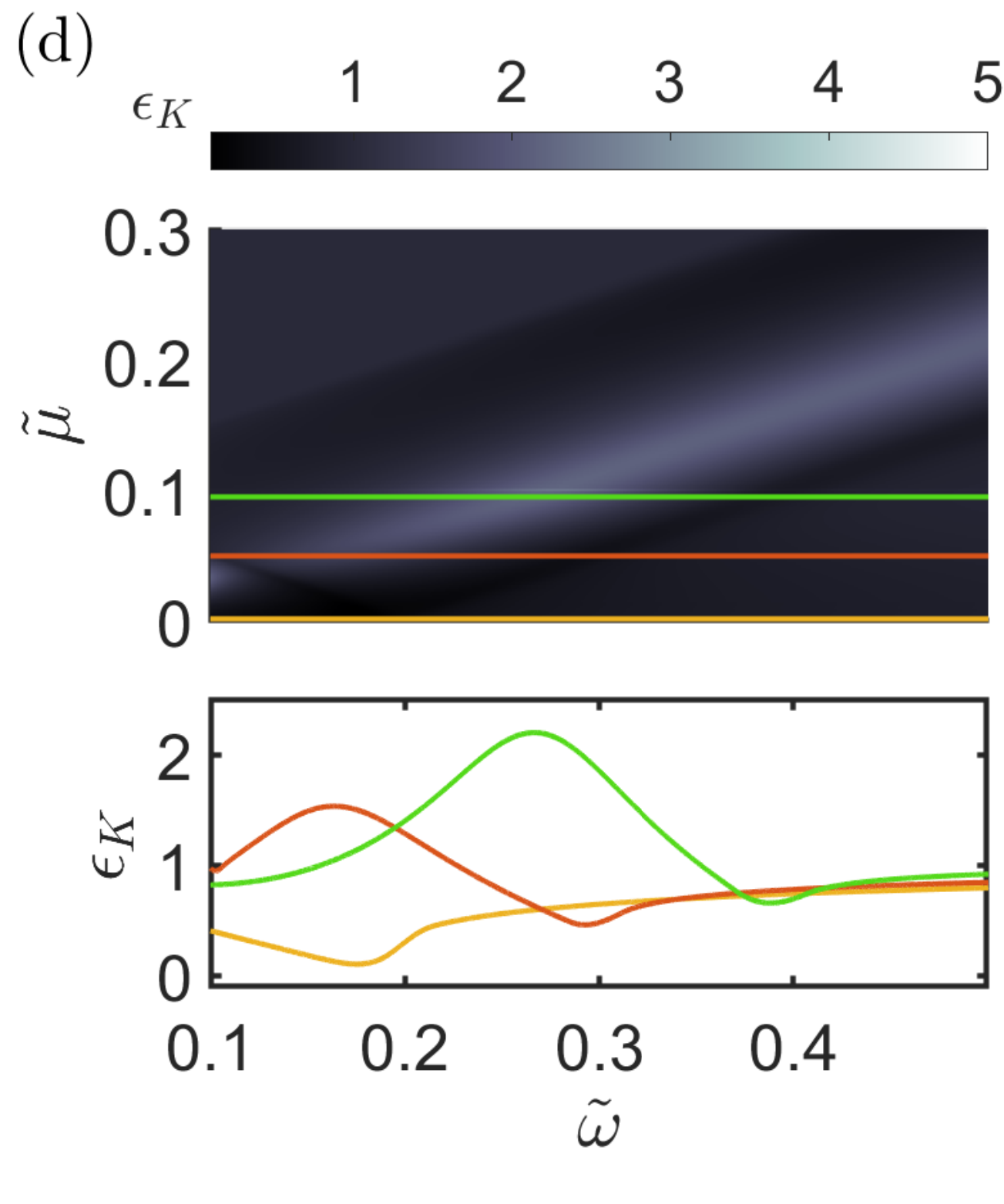}
	\caption{\emph{(a,b):} Kerr angle $ \theta_K $ for a thin film. \emph{(c,d): } Kerr ellipticity $ \epsilon_{K} $ for a thin film. For (a) and (c) we fix the chemical potential at $\tilde{\mu} = 0$ and vary the tilt velocity. The sharp peak dip corresponds to $ \tilde{\omega} = 2\tilde{u} $. For (b) and (d) we fix the tilt velocity at $ \tilde{u} = 0.1 $ and vary the chemical potential. }
	\label{Fig:KerrT}
\end{figure}

The boundary conditions applied to Maxwell's equations on the two sides of a thin metallic film state that the electric field components parallel to the surface are continuous across the boundary, while there is a discontinuity in the magnetic field equal to the generated current \cite{Farias20}. In mathematical terms they are written as \cite{jackson07l}

\begin{align}
& \mathbf{E}_{1}^{\parallel} =  \mathbf{E}_{2}^{\parallel}, \label{Eq:BoundaryMaxwell2} \\
& \hat{\mathbf{n}}_{12}\times\left (\frac{\mathbf{B}_{1}^{\parallel}}{\mu_{1}} - \frac{\mathbf{B}_{2}^{\parallel}}{\mu_{2}}\right ) = \frac{4\pi}{c}\mathbf{J},\label{Eq:BoundaryMaxwell3}
\end{align}
Here, $\mathbf{E}_i^{\parallel}$ (for the two regions $i = 1,2$) are the components of the electric field vector inside the surface plane above and below the film, respectively. Moreover, $ \hat{\mathbf{n}}_{12} $ is the surface normal vector pointing from medium $ 1 $ to medium $ 2 $. The magnetic field is given by $\mathbf{B} = \frac{c}{\omega}\mathbf{k}\times\mathbf{E} $ and $ \mu_{i} $ is the magnetic permeability of medium $ i $. Since we consider vacuum on both sides of the film, we set $ \mu_{i} = 1 $.
Finally the current density is given by $\mathbf{J} = \sigma^{S}\mathbf{E}$, where the surface conductivity tensor can be approximated by the bulk conductivity and the film thickness, $ d $ such that $ \sigma_{ij}^{S} = d\sigma_{ij} $ \cite{Kargarian2015}.

We consider an incoming wave
\begin{equation}
\mathbf{E}_{0} = \frac{E_{0}}{2}\left(\hat{e}_{R} + \hat{e}_{L} \right)e^{i(k_{z}z-\omega t)},
\label{Eq:IncidentWave}
\end{equation}
which describes an electromagnetic wave propagating along the $k_{z}$ direction and linearly polarized along the $x$ axis (here decomposed into two circularly polarized waves with left- and right-handed polarization) and with an amplitude $E_{0}$. In this case, the reflected (subscript $r$) and transmitted ($t$) waves, respectively, are given by
\begin{align}
\mathbf{E}_{r} &= \left(E_{r}^{R}\hat{e}^{\prime}_{R} + E_{r}^{L}\hat{e}^{\prime}_{L}\right)e^{-i(k_{z}z + \omega t)}, \label{Eq:ReflectedWave} \\
\mathbf{E}_{t} &= \left(E_{t}^{R}\hat{e}_{R} + E_{t}^{L}\hat{e}_{L}\right)e^{i(k_{z}z - \omega t)},
\end{align}
where $ \hat{e}^{\prime}_{R,L} = \mathbf{\hat{x}} \pm i \mathbf{\hat{y}}$ is the circularly polarized basis for the reflected light. Calculating the corresponding magnetic fields, inserting the fields into the boundary conditions \eqref{Eq:BoundaryMaxwell2} and \eqref{Eq:BoundaryMaxwell3} and using that $\sigma_{yx} = -\sigma_{xy}$ we obtain a system of equations which we can solve for $E_{r}^{R}$ and $E_{r}^{L}$. We obtain
\begin{align}
E_{r}^{R,L} = \frac{E_{0}}{C}\left[\left(2\kappa - \sigma_{2}^{\pm}\right) \sigma_{xx}^{S} \mp i\left(2\kappa - \sigma_{1}^{\pm}\right)\sigma_{xy}^{S} \right],
\label{Eq:reflection_thinfilm}
\end{align}
where
\begin{align}
    C
&=
    \left(2\kappa - \sigma_{1}^{-}\right)\left(2\kappa - \sigma_{2}^{+}\right) + \left(2\kappa - \sigma_{1}^{+}\right)\left(2 \kappa- \sigma_{2}^{-}\right), \notag \\ \sigma_{1}^{\pm}
&=
    \sigma_{xx}^{S} \pm i\sigma_{xy}^{S}, \notag \\
    \sigma_{2}^{\pm}
&=
    \sigma_{yy}^{S} \pm i\sigma_{xy}^{S},
\end{align}
and $ \kappa = 1/(4\pi\alpha)$ with the fine structure constant $ \alpha \approx 1/137 $. The upper signs in Eq.~\eqref{Eq:reflection_thinfilm} correspond to the right handed polarization and the lower signs to the left handed one. To simplify the result, we make the assumption that the film thickness is much smaller than the wavelength of the incident light. The Kerr angle and ellipticity can now be calculated using Eq.~\eqref{Eq:KerrAngle}. We plot the result for a tilt angle $ \theta = \pi/4 $ in Fig.~\ref{Fig:KerrT}, and more general angles are discussed in Sec.~\ref{Sec:VaryTilt}.

At zero chemical potential, Figs.~\ref{Fig:KerrT}(a,c) show a pronounced dip in the Kerr angle and ellipticity at the frequency $ \tilde{\omega} = 2\tilde{u} $. This is the point where the real part of the transverse conductivity vanishes and the imaginary part reaches its maximum. Hence, a finite tilt of the nodal line has a strong effect on the Kerr signal. By increasing the chemical potential, see Figs.~\ref{Fig:KerrT}(b,d), we notice that the Kerr angle is reduced but still remains large. It reaches its largest values between $ \tilde{\omega} \approx 2\tilde{\mu} + \tilde{u}/2 $ and $ \tilde{\omega} \approx 2\tilde{\mu} + 2\tilde{u} $. For $ \tilde{\mu} > \tilde{u} $, this is proceeded by a region, defined by $ 2\tilde{\mu} - 3\tilde{u}/2 \lessapprox \tilde{\omega} \lessapprox 2\tilde{\mu} + \tilde{u}/2 $, where it takes negative values on the same order of magnitude.

\subsection{Bulk material}

\begin{figure}[t]
	\centering
	\includegraphics[width=0.23\textwidth]{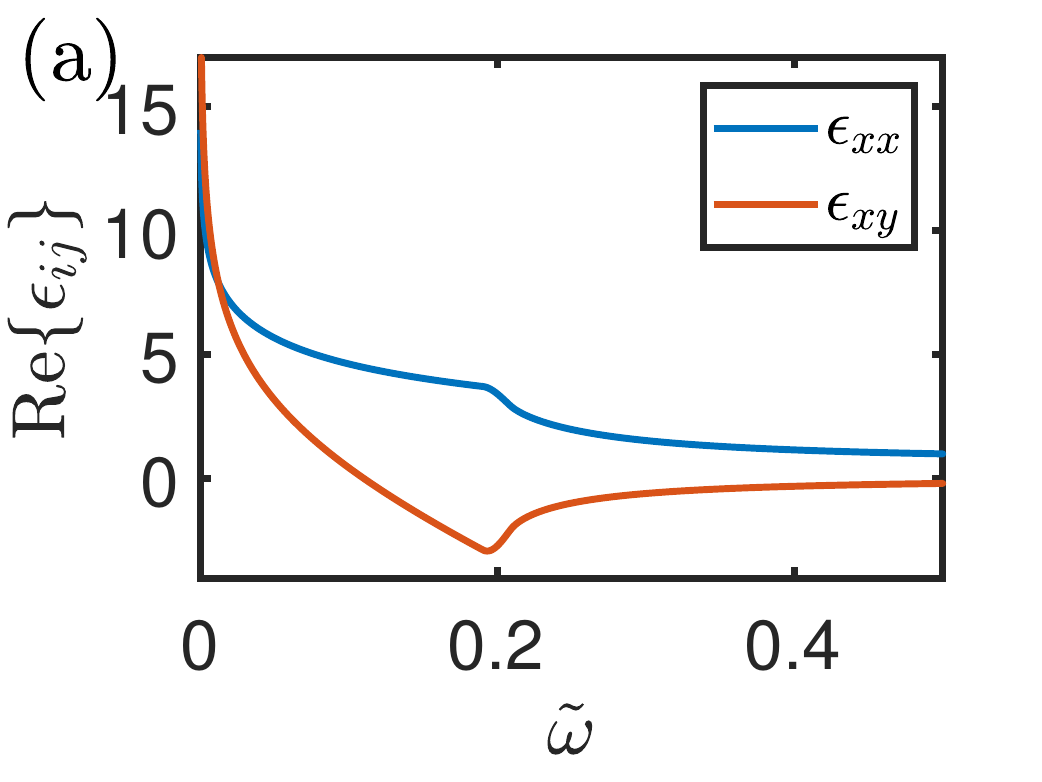}
	\includegraphics[width=0.23\textwidth]{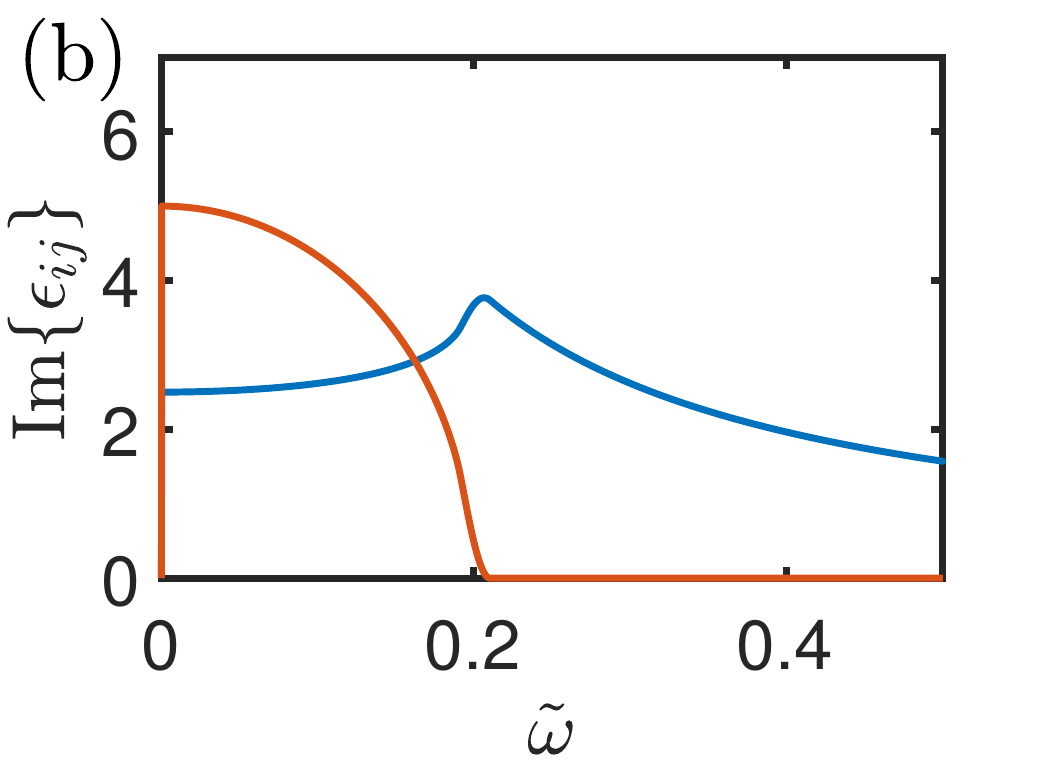}
	\caption{Real (a) and imaginary (b) parts of the components of the permittivity for $ \theta = \pi/4 $, $ \tilde{u} = 0.1 $, $ \tilde{\mu} = 0 $ and $ \epsilon_{b} = 1 $.}
	\label{Fig:DiElec}
\end{figure}

Next, we consider the Kerr reflection on the surface of a bulk material. Compared to the thin film we now have to consider the propagation of the electromagnetic waves inside the material. Hence, it is necessary to find the allowed wave vectors inside the bulk material and this is done by solving the electromagnetic wave equation obtained from Maxwell's equations. The Maxwell-Faraday and Ampere-Maxwell equations are respectively given by
\begin{align}
	\nabla \times \mathbf{E} &= -\frac{1}{c}\frac{\partial \mathbf{B}}{\partial t}, \label{Eq:MF} \\
	\nabla\times \mathbf{H} &= \frac{1}{c}\frac{\partial \mathbf{D}}{\partial t} + \frac{4\pi}{c}\mathbf{J}. \label{Eq:AM}
\end{align}
We are assuming that the material is non-magnetic and hence $ \mathbf{B} = \mathbf{H} $. The constitutive relations further tells us that $ \mathbf{D} = \epsilon_{b}\mathbf{E} $, where $ \epsilon_{b} $ is the static permittivity of the nodal loop material. For the following investigation we select $ \epsilon_{b} = 10 $ as a realistic value. We take the curl of Eq.~\eqref{Eq:MF} and insert Eq.~\eqref{Eq:AM} along with the constitutive relations and Ohm's law $ \mathbf{J} = \sigma\mathbf{E} $ into the obtained expression. The result is,
\begin{equation}
\nabla\times\left(\nabla\times\mathbf{E}\right) = -\frac{1}{c^{2}}\left[ \epsilon_{b}\frac{\partial^{2}\mathbf{E}}{\partial t^{2}} + 4\pi\sigma\frac{\partial \mathbf{E}}{\partial t}\right].	
\end{equation}
Expanding the curl on the left hand side and performing a Fourier transform, we obtain
\begin{equation}
(\mathbf{k} \cdot \mathbf{k})\mathbf{E}(\mathbf{k},\omega) - \mathbf{k}\left[\mathbf{k}\cdot\mathbf{E}(\mathbf{k},\omega)\right] = \epsilon\frac{\omega^{2}}{c^{2}}\mathbf{E}(\mathbf{k},\omega).
\label{Eq:WaveEquationFourier}
\end{equation}
Here we have introduced the permittivity tensor
\begin{equation}
\epsilon(\omega) = \begin{pmatrix}
\epsilon_{xx} & \epsilon_{xy} & 0 \\
\epsilon_{yx} & \epsilon_{yy} & 0 \\
 0 & 0 & \epsilon_{zz}
\end{pmatrix}.
\end{equation}
and it is fully determined by the following relation to the conductivity tensor: $\epsilon_{ij} = \delta_{ij}\epsilon_{b} + \frac{4\pi i}{\omega}\sigma_{ij}$
The components of the permittivity tensor are plotted in Fig.~\ref{Fig:DiElec} for a nodal loop with $\theta = \pi/4$.

\begin{figure}[t]
	\includegraphics[width=0.23\textwidth]{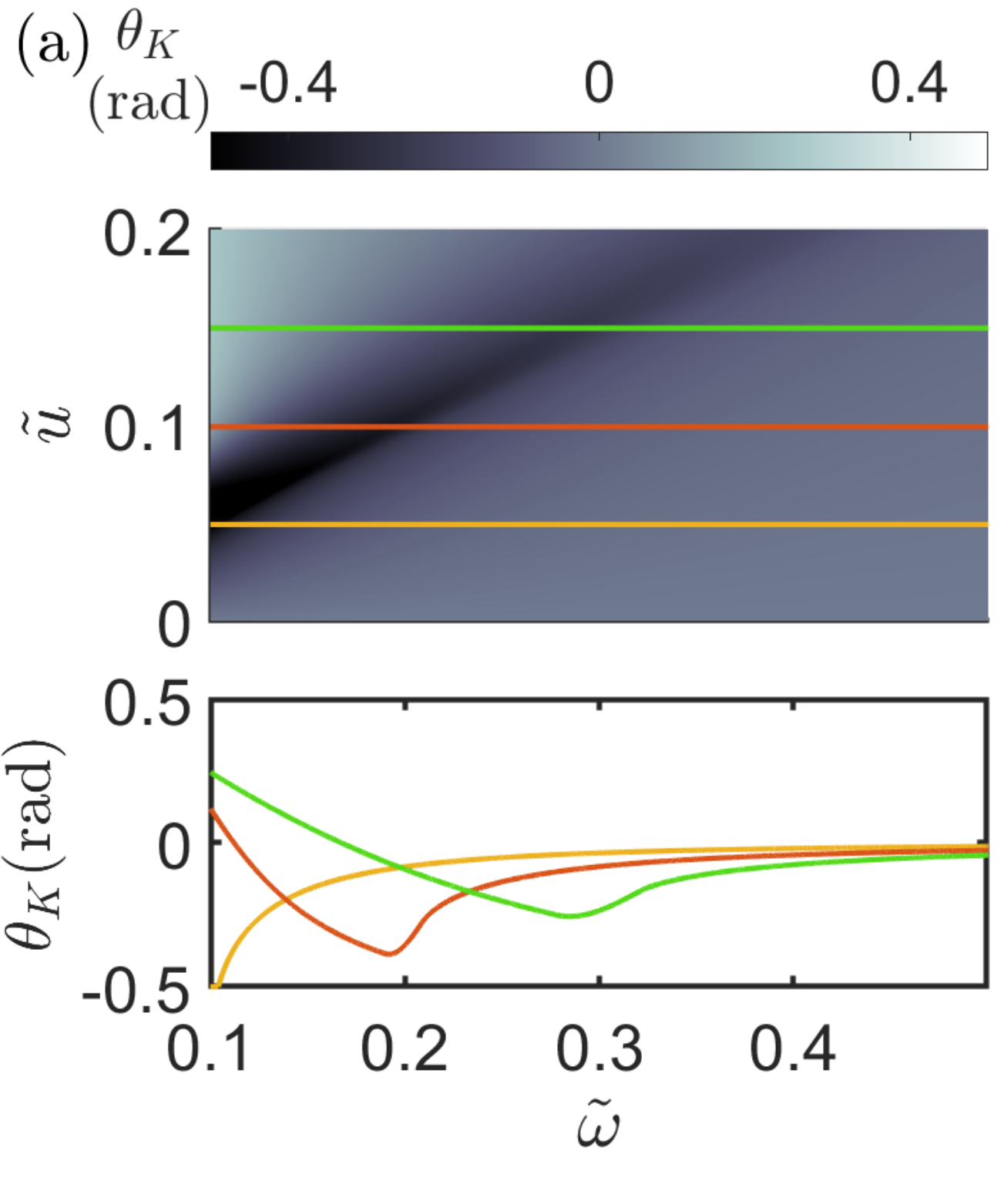}
	\includegraphics[width=0.23\textwidth]{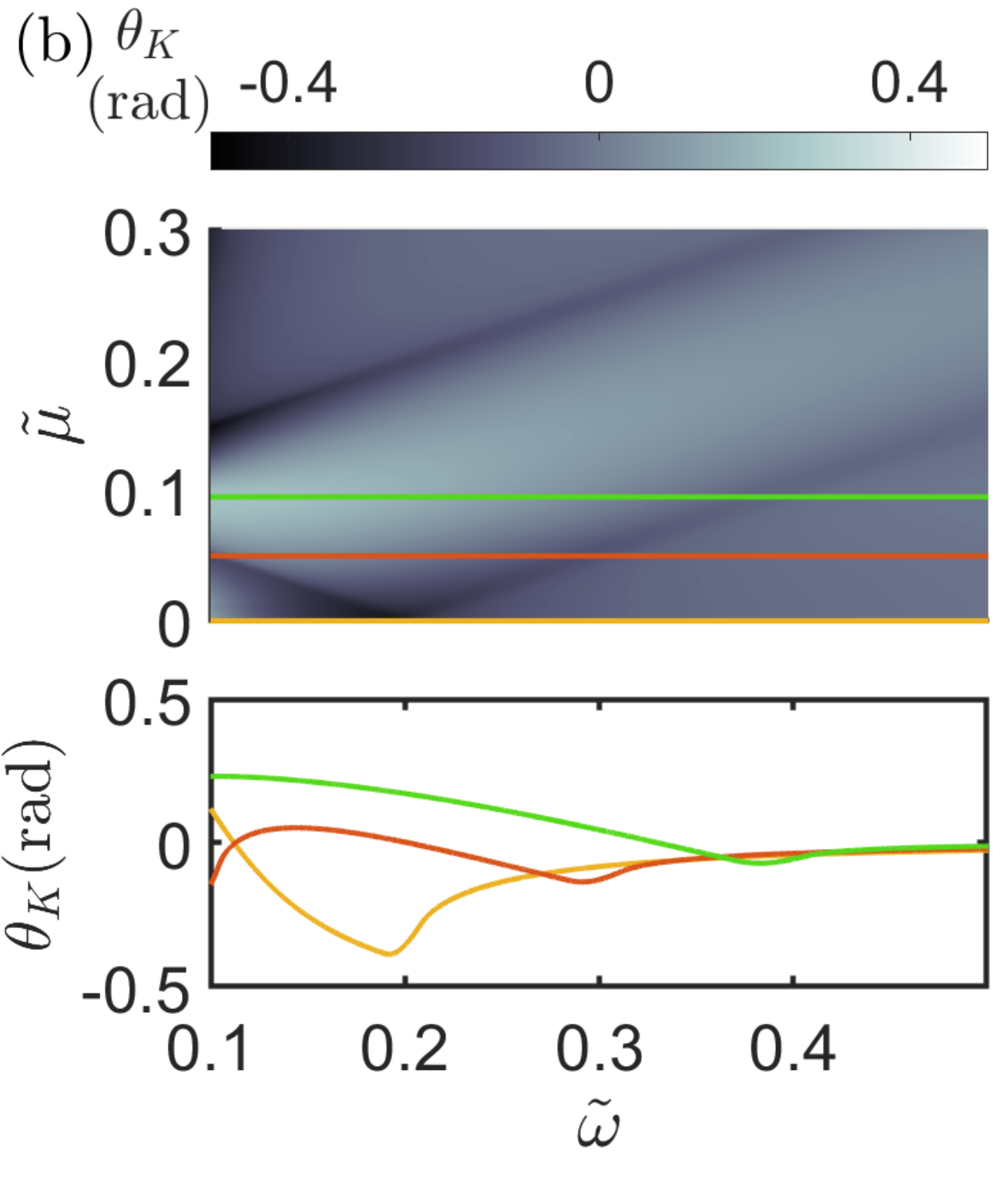}
	\includegraphics[width=0.23\textwidth]{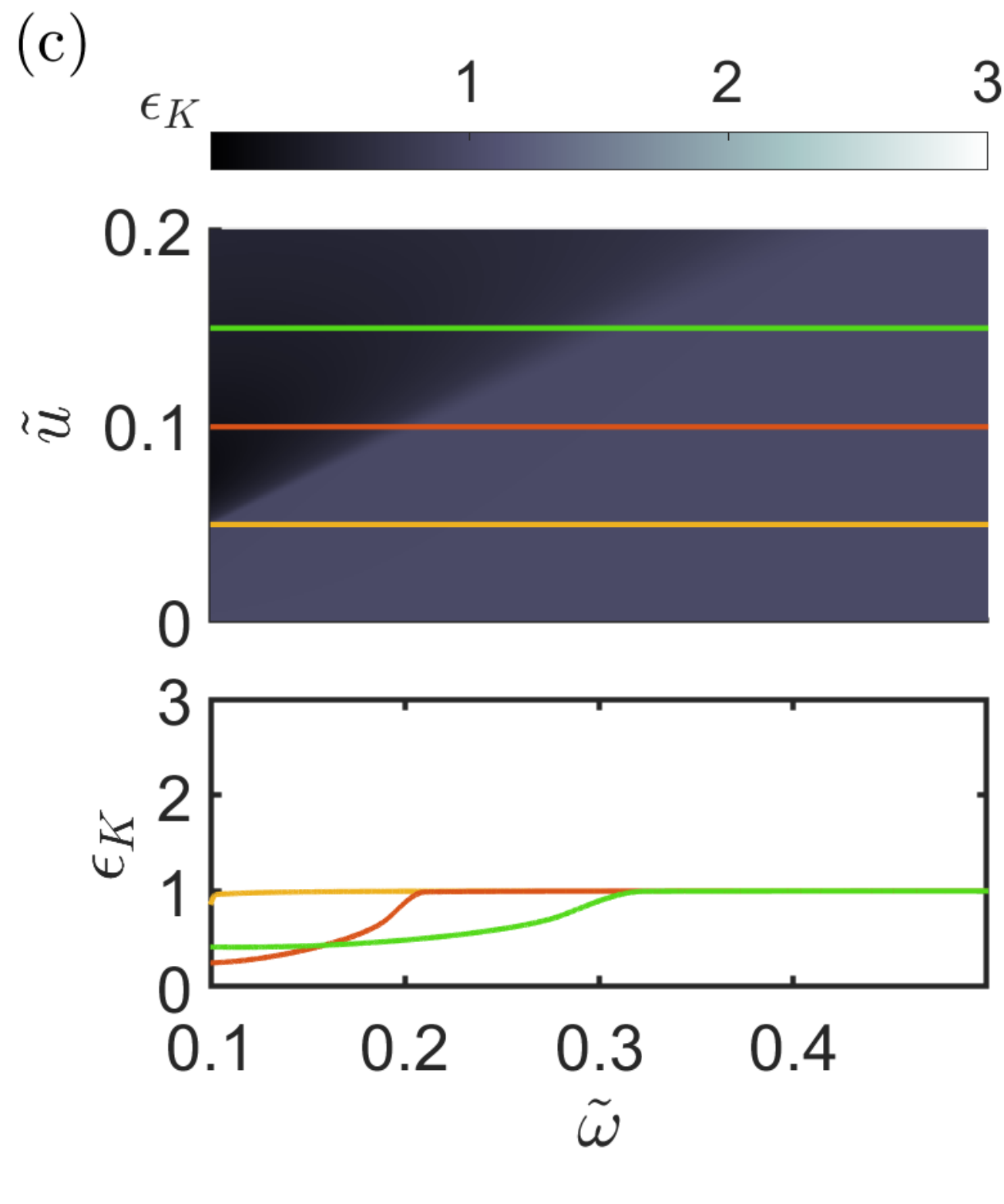}
	\includegraphics[width=0.23\textwidth]{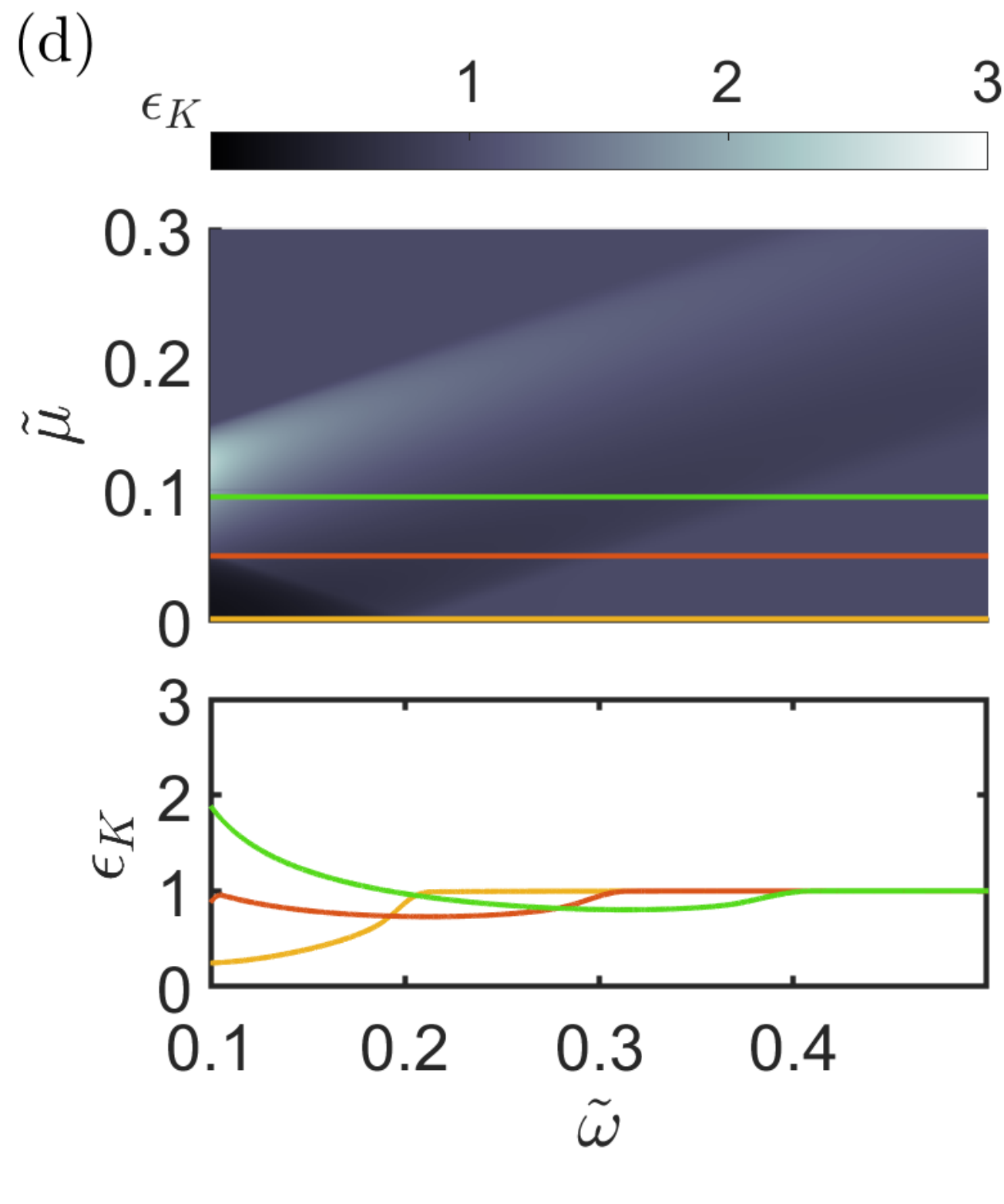}
	\caption{(\emph{a,b}) Kerr angle and (\emph{c,d}) Kerr ellipticity for the bulk material with light incident of the $ x-y $ surface. For (a) and (c) we fix the chemical potential, $\tilde{\mu} = 0$ and vary the tilt velocity. The sharp dip corresponds to $ \omega = 2\tilde{u} $. For (b) and (d) we fix the tilt velocity, $ \tilde{u} = 0.1 $ and vary the chemical potential.}
	\label{Fig:KerrB}
\end{figure}

\subsubsection{Propagation along the $ k_{z} $-direction}

We start again with an incoming beam propagating along the $k_{z}$ axis. The transmitted field can be written as $\mathbf{E}_{t}(z,t) = \mathbf{E}_{t}(k_{z}^{\prime},\omega)e^{ik_{z}^{\prime}z - i\omega t}$, where $k_z'$ is the wave vector inside the bulk material. Inserting this into Eq.~\eqref{Eq:WaveEquationFourier} it follows that for nontrivial solutions the following determinant vanishes,
\begin{equation}
	\begin{vmatrix}
		k'^{2}_{z} + \frac{\omega^{2}}{c^{2}}\epsilon_{xx} & \frac{\omega^{2}}{c^{2}}\epsilon_{xy} \\ -\frac{\omega^{2}}{c^{2}}\epsilon_{xy} & k'^{2}_{z} + \frac{\omega^{2}}{c^{2}}\epsilon_{yy}
	\end{vmatrix} = 0.
\end{equation}
From here we find that the allowed wave vectors inside the material are given by
\begin{align}
k'^{2}_{z} &= \frac{\omega^{2}}{2c^{2}}(\epsilon_{xx} + \epsilon_{yy}) \nonumber \\ &\hspace{5pt} \pm\frac{\omega^{2}}{c^{2}} \sqrt{\frac{1}{4}\left(\epsilon_{xx} + \epsilon_{yy}\right)^{2} - \epsilon_{xx}\epsilon_{yy} - \epsilon_{xy}^{2}} \equiv k_{\pm}^{2}.
\label{Eq:kpm}
\end{align}
We further find the basis vectors for the allowed electromagnetic fields inside the material. They are given by
\begin{align}
	\mathbf{e}_{t}^{\pm} = c_{1}^{\pm}\mathbf{\hat{x}} + c_{2}^{\pm}\mathbf{\hat{y}},
\end{align}
with the coefficients $c_{1}^{\pm} = |\epsilon_{xy}| (|\epsilon_{\pm} - \epsilon_{xx}|^{2} + |\epsilon_{xy} |^{2})^{-1/2} $, $c_{2}^{\pm} = \mathrm{sgn}(\epsilon_{xy})(\epsilon_{\pm} - \epsilon_{xx}) (|\epsilon_{\pm} - \epsilon_{xx}|^{2} + |\epsilon_{xy} |^{2})^{-1/2} $ and $\epsilon_{\pm} = c^{2}k_{\pm}^{2}/\omega^{2}$. The total transmitted wave $\mathbf{E}_{t}$ can be written as the linear combination $\mathbf{E}_{t} = E^{+}\mathbf{e}_{t}^{+} + E^{-}\mathbf{e}_{t}^{-}$. The boundary conditions are now given by
\begin{align}
\mathbf{E}_{1}^{\parallel} &= \mathbf{E}_{2}^{\parallel}, \label{Eq:BC3} \\
\frac{1}{\mu_{1}}\mathbf{B}_{1}^{\parallel} &= \frac{1}{\mu_{2}}\mathbf{B}_{2}^{\parallel}\label{Eq:BC4},
\end{align}
where again the $ \parallel $ superscript denotes the field components parallel to the surface. We take the incident and reflected waves, Eqs.~\eqref{Eq:IncidentWave} and \eqref{Eq:ReflectedWave}, and insert these along with the expression for the transmitted wave into the boundary conditions, Eqs.~\eqref{Eq:BC3}, and \eqref{Eq:BC4}. By solving the system of equations we obtain the reflected fields,
\begin{align}
E_{r}^{R,L} &= \frac{1}{4}\left ( 1 - \frac{c}{\omega}k_{+} \right )\left ( c_{1}^{+} \mp ic_{2}^{+} \right ) E_{+} \nonumber \\ &  + \frac{1}{4}\left ( 1 - \frac{c}{\omega}k_{-} \right )\left ( c_{1}^{-} \mp ic_{2}^{-} \right ) E_{-},
\end{align}
where
\begin{equation}
E_{\pm} = \frac{4E_{0}}{\left ( 1 + \frac{c}{\omega}k_{\pm} \right )\left ( c_{1}^{\pm} - \frac{c_{2}^{\pm}}{c_{2}^{\mp}} c_{1}^{\mp} \right )},
\end{equation}
where $ k_{\pm} $ is defined in Eq.~\eqref{Eq:kpm}. These allow us to calculate the Kerr angle and Kerr ellipticity.

The results shown in Fig.~\ref{Fig:KerrB} show similar patterns as those for the thin-film geometry, especially for $ \tilde{\mu} = 0 $. The main difference in this case lies in the amplitude of the Kerr angle. In comparison to the thin film, the amplitude is smaller and is further reduced when the static permittivity $ \epsilon_{b} $ is increased. The Kerr rotation should still be experimentally detectable since Kerr angles of order of $ 10^{-9} $~radians have been measured \cite{Xia06}. We further note that for larger frequencies $ \epsilon_{K} = 1 $. This is a consequence of the fact that excitations become available along the whole nodal line and $ \mathrm{Re}\lbrace\sigma_{xy}\rbrace = 0 $.

When the chemical potential is increased, Fig.~\ref{Fig:KerrB}(b), the dip at $ \tilde{\omega} = 2\tilde{u} $ splits into two dips centered around $ \tilde{\omega} = 2\tilde{u}\pm 2\tilde{\mu} $. As $ \tilde{\mu} $ becomes greater than $ \tilde{u} $ the peaks are centered around $ \tilde{\omega} = 2\tilde{\mu} $ and $ \tilde{\omega} = 2\tilde{u} + 2\tilde{\mu} $.
As the chemical potential becomes larger than the tilt velocity the system becomes Pauli blocked. This is directly observed in the ellipticity, Fig.~\ref{Fig:KerrB}(d), which equals 1 in the region where the system is Pauli blocked (upper left region of the colormap).

\subsubsection{Propagation along the $ k_{x} $-direction}

Next, we will consider an incoming wave travelling along the $k_x$ direction and impinging on the $ y$-$z $ surface. We return to Eq.~\eqref{Eq:WaveEquationFourier} and solve it for a transmitted field given by $\mathbf{E}_{t}(x,t) = \mathbf{E}_{t}(k_{x}^{\prime},\omega)e^{ik_{x}^{\prime}x - i\omega t}$, where $k_x'$ is the allowed wave vector inside the bulk material given an incoming field propagating along the $ k_{x} $-direction. Inserting the field into Eq.~\eqref{Eq:WaveEquationFourier} we find the allowed wave vectors by solving the following determinant
\begin{equation}
	\begin{vmatrix}
		\frac{\omega^{2}}{c^{2}}\epsilon_{xx} & \frac{\omega^{2}}{c^{2}}\epsilon_{xy} & 0 \\
		-\frac{\omega^{2}}{c^{2}}\epsilon_{xy} & -\left(k_{x}^{\prime}\right )^{2} + \frac{\omega^{2}}{c^{2}}\epsilon_{yy} & 0 \\
		0 & 0 & -\left(k_{x}^{\prime}\right)^{2} + \frac{\omega^{2}}{c^{2}}\epsilon_{zz}
	\end{vmatrix} = 0
\end{equation}
The solution gives wave-vectors of two types,
\begin{align}
	\left(k_{x}^{\prime}\right)^{2} &= \frac{\omega^{2}}{c^{2}}\epsilon_{zz} \equiv k_{1}^{2}, \\
	\left(k_{x}^{\prime}\right)^{2} &= \frac{\omega^{2}}{c^{2}}\left(\epsilon_{yy} - \frac{\epsilon_{xy}^{2}}{\epsilon_{xx}}\right) \equiv k_{2}^{2}.
\end{align}
As for the previous case we find the basis for the allowed electromagnetic fields. In this case we have to use a different basis due to the different solutions for the allowed wave vectors. We then have
\begin{align}
	\mathbf{e}^{1}_{t} &= \hat{\mathbf{z}}, \\
	\mathbf{e}_{t}^{2} &= a\frac{\epsilon_{xy}}{\epsilon_{xx}}\hat{\mathbf{x}} - a\hat{\mathbf{y}},
\end{align}
where $ a = (1 + |\epsilon_{xy}/\epsilon_{xx}|^{2})^{-1/2}$. The total transmitted electric field can then be written as $ \mathbf{E}_{t}^{\prime} = E^{2}\mathbf{e}_{t}^{2} + E^{1}\mathbf{e}_{t}^{1} $. Using this field and combining it with the boundary conditions given by Eqs.~\eqref{Eq:BC3} and \eqref{Eq:BC4} we obtain the reflected fields
\begin{align}
	E_{r}^{R,L} = \frac{1}{2}\left[ \frac{1 - \frac{c}{\omega}k_{2}}{1 + \frac{c}{\omega}k_{2}}
	a E_{0}^{y} \mp i\frac{1 - \frac{c}{\omega}k_{1}}{1 + \frac{c}{\omega}k_{1} }
	E_{0}^{z} \right]
	\label{Eq:kx_Efield}
\end{align}
It can be seen that if the incident linearly polarized electric field only has either a $ \hat{\mathbf{y}} $ or $ \hat{\mathbf{z}} $ component then $ E_{r}^{R} = E_{r}^{L} $ and hence neither a Kerr angle nor an ellipticity should be observed. Furthermore, the reflected field remains linearly polarized. Hence, for the ensuing discussion we focus on the case when $ E_{0}^{y} = E_{0}^{z} = 1/\sqrt{2} $.

\begin{figure}[t]
	\includegraphics[width=0.23\textwidth]{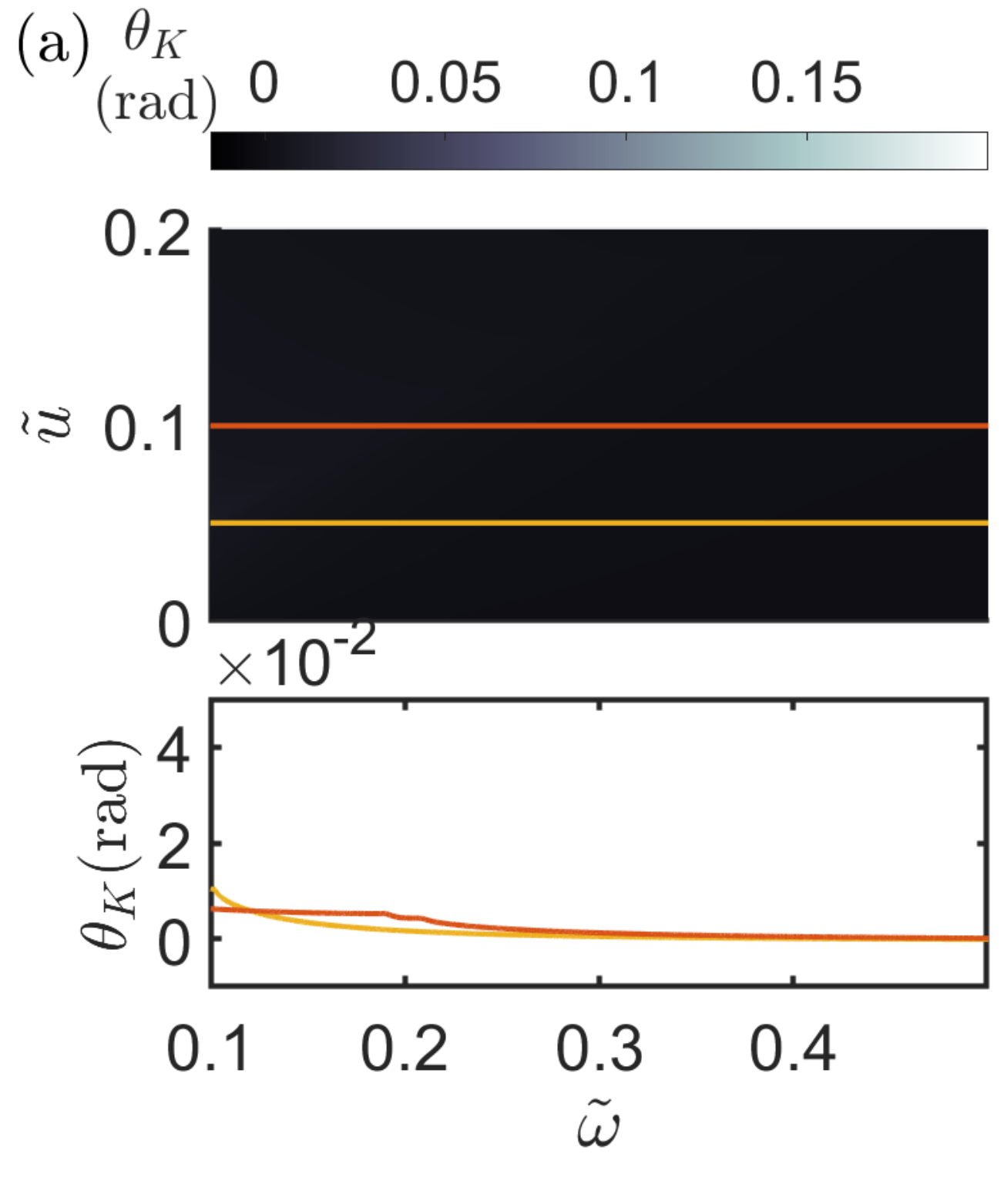}
	\includegraphics[width=0.23\textwidth]{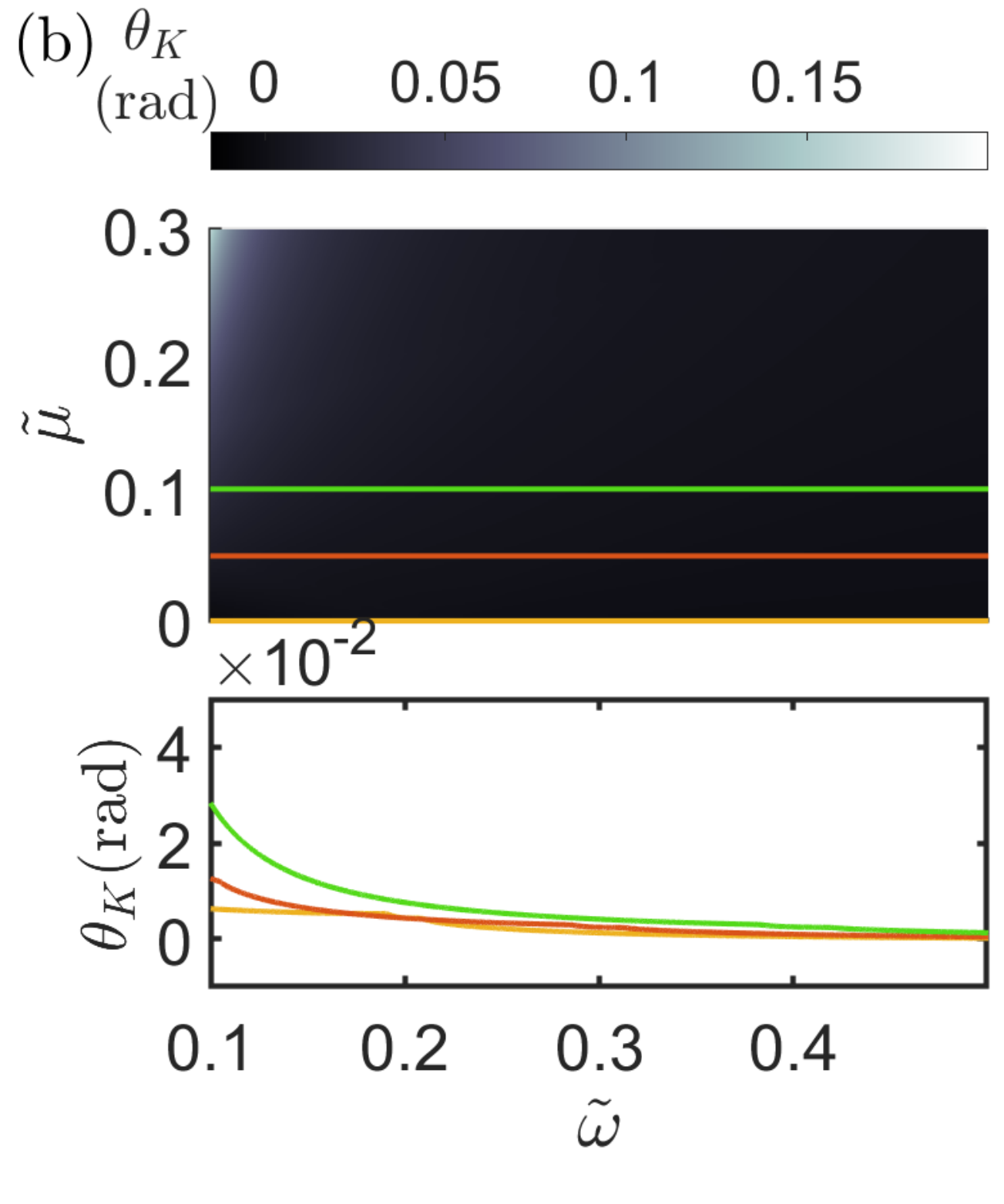}
	\includegraphics[width=0.23\textwidth]{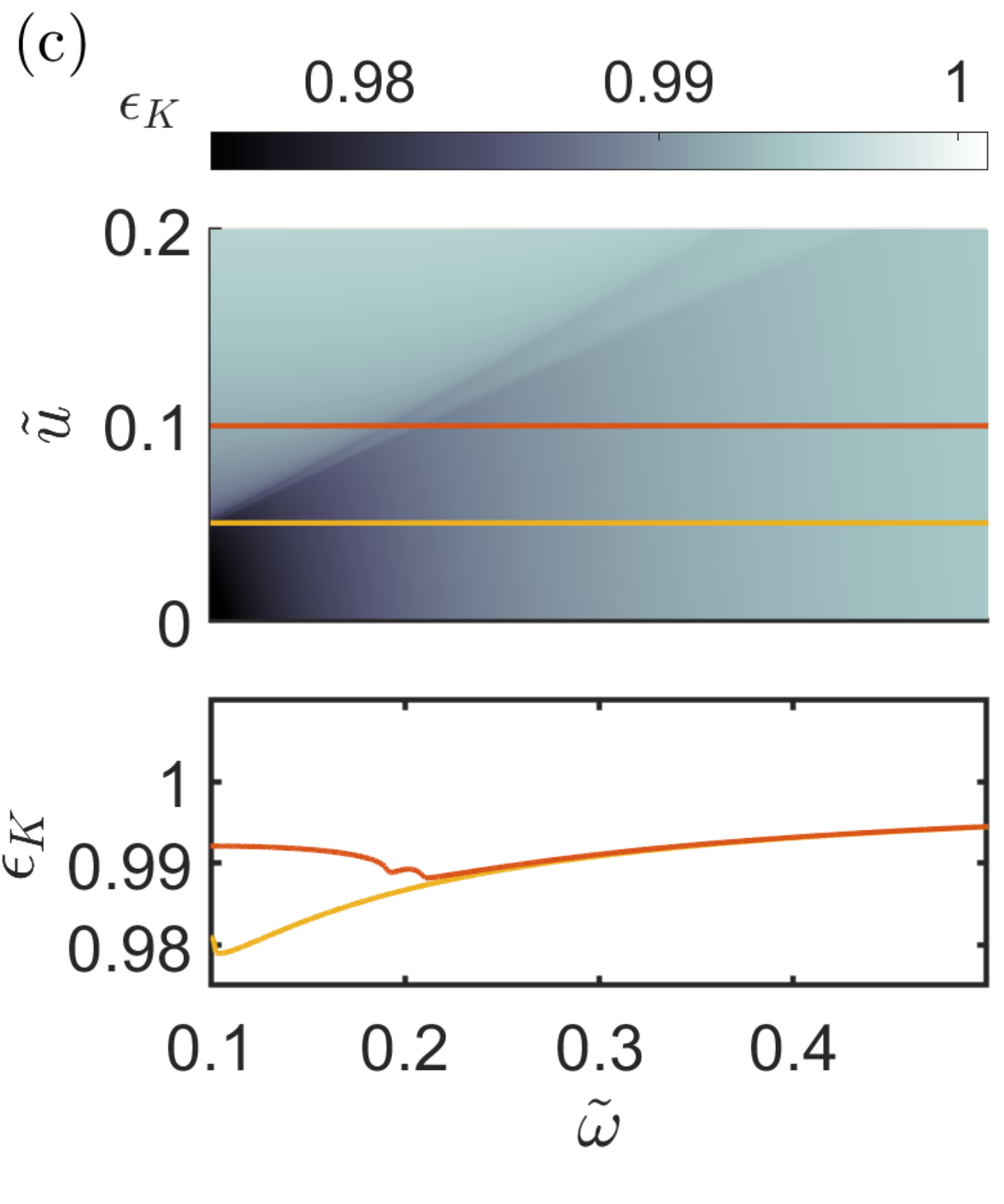}
	\includegraphics[width=0.23\textwidth]{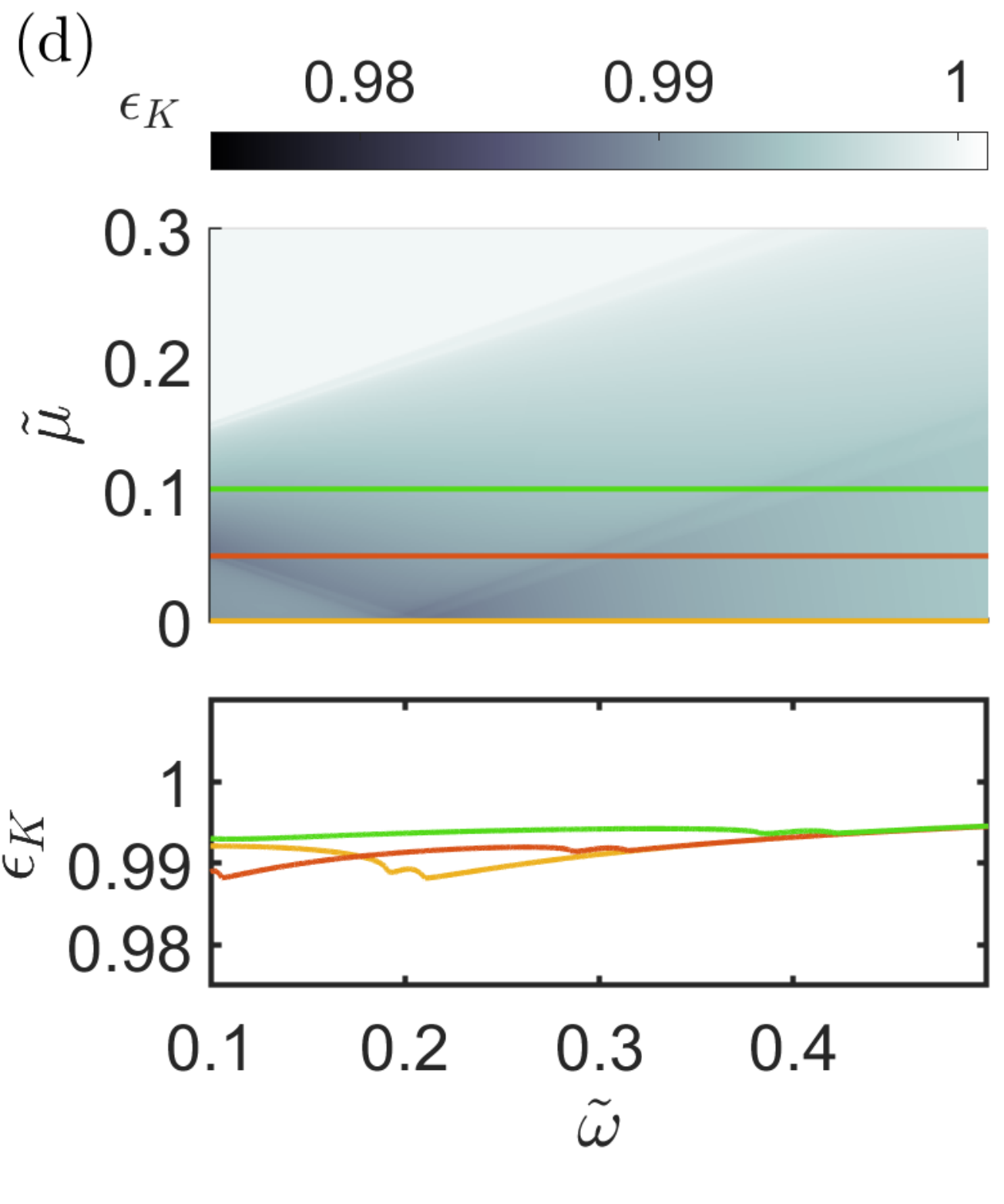}
	\caption{(\emph{a,b}) $ \theta_K $ for bulk material and (\emph{c,d}) $ \epsilon_{K} $ for bulk material and for a beam incident on the $ y-z $ surface. For (a) and (c) we fix the chemical potential, $\tilde{\mu} = 0$ and vary the tilt velocity. For (b) and (d) we fix the tilt velocity, $ \tilde{u} = 0.1 $ and vary the chemical potential.}
	\label{Fig:KerrBYZ}
\end{figure}

Using Eqs.~\eqref{Eq:KerrAngle} and \eqref{Eq:kx_Efield} we plot the Kerr rotation for incidence along the $k_x$ axis in Fig.~\ref{Fig:KerrBYZ}. For $ \mu = 0 $ (Fig.~\ref{Fig:KerrBYZ}~(a) and (c)) we observe quite different features as we vary the tilt velocity, compared to incidence along the $z$ axis. The Kerr angle is an order of magnitude smaller and the ellipticity remains close to $ 1 $ for all frequencies. This is due to the fact that $ \epsilon_{yz} = 0 $, which otherwise would strongly contribute to the Kerr angle and the ellipticity. Furthermore, the contribution from $ \epsilon_{xy} $ is strongly suppressed by $ \epsilon_{xx} $, see Eq.~\eqref{Eq:kx_Efield}.

As the chemical potential is increased (see Fig.~\ref{Fig:KerrBYZ}~(b) and (d)) the features of the Kerr angle are similar to when $ \mu = 0 $. The amplitude however increases for small $ \tilde{\omega} $. This is due to the increasing amplitude of the intraband transitions. The ellipticity on the other hand is shifted towards $ \epsilon_{K} = 1 $. As for incidence on the $ x$-$y $ surface, this is because the system becomes Pauli blocked.

\subsection{Varying the tilt angle} \label{Sec:VaryTilt}

\begin{figure}[t]
	\includegraphics[width=0.23\textwidth]{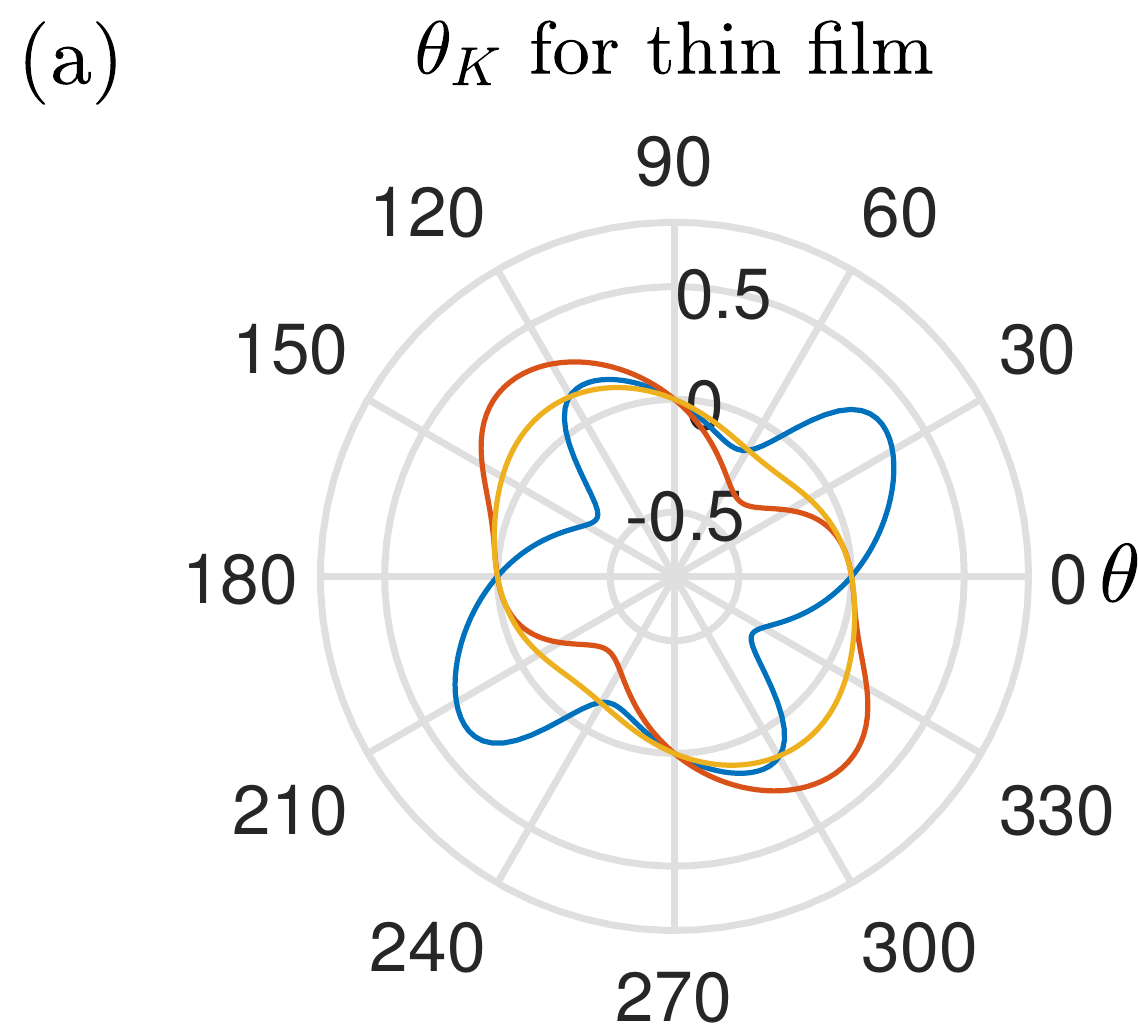}
	\includegraphics[width=0.23\textwidth]{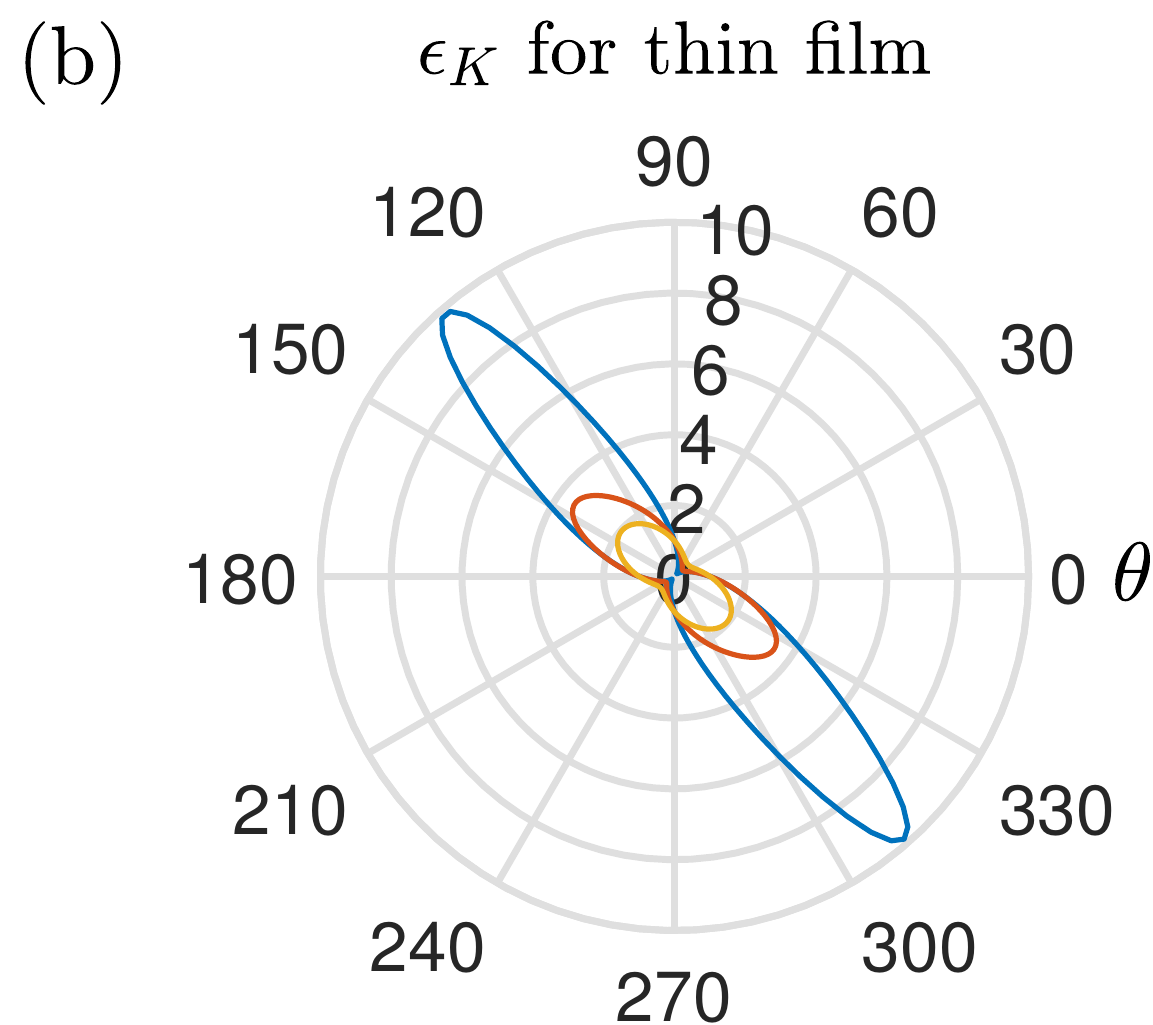}
	\includegraphics[width=0.23\textwidth]{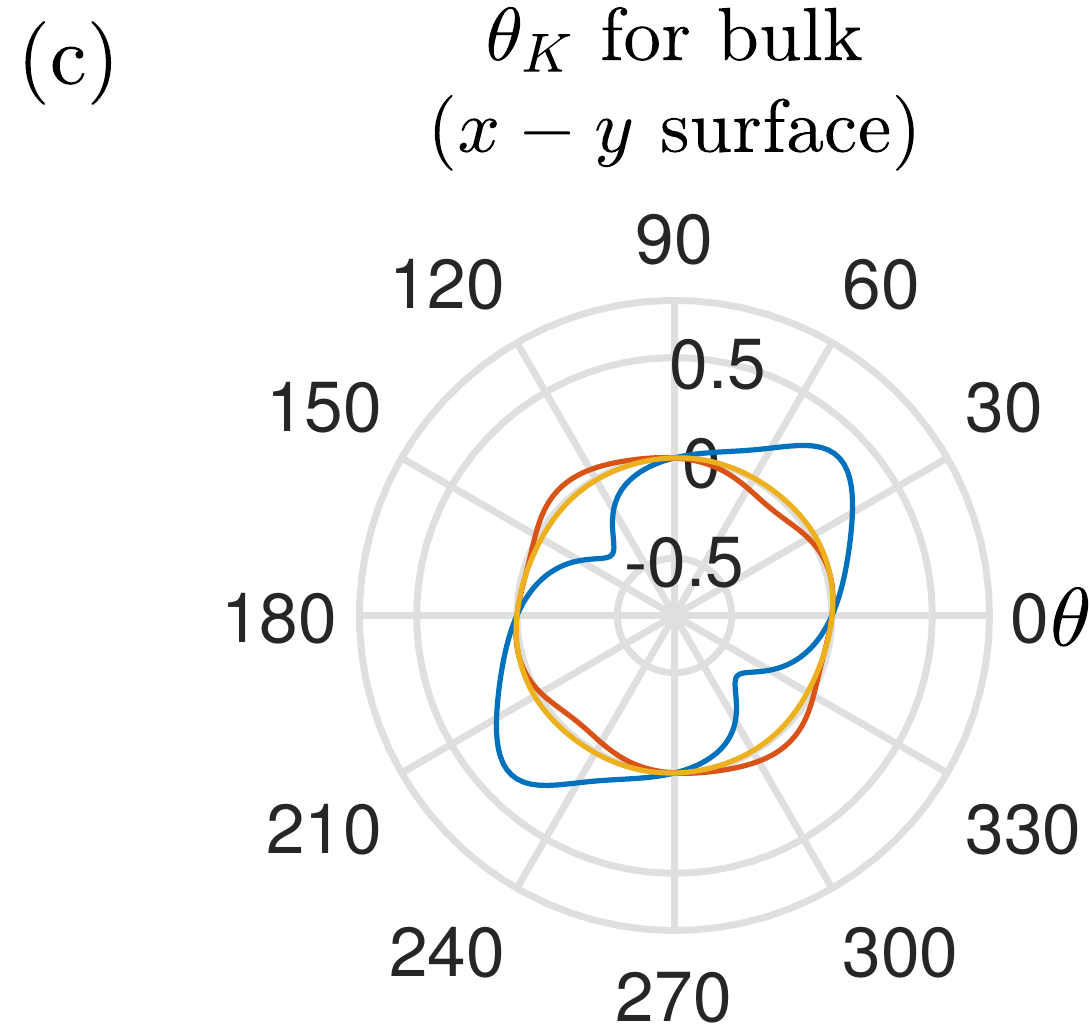}
	\includegraphics[width=0.23\textwidth]{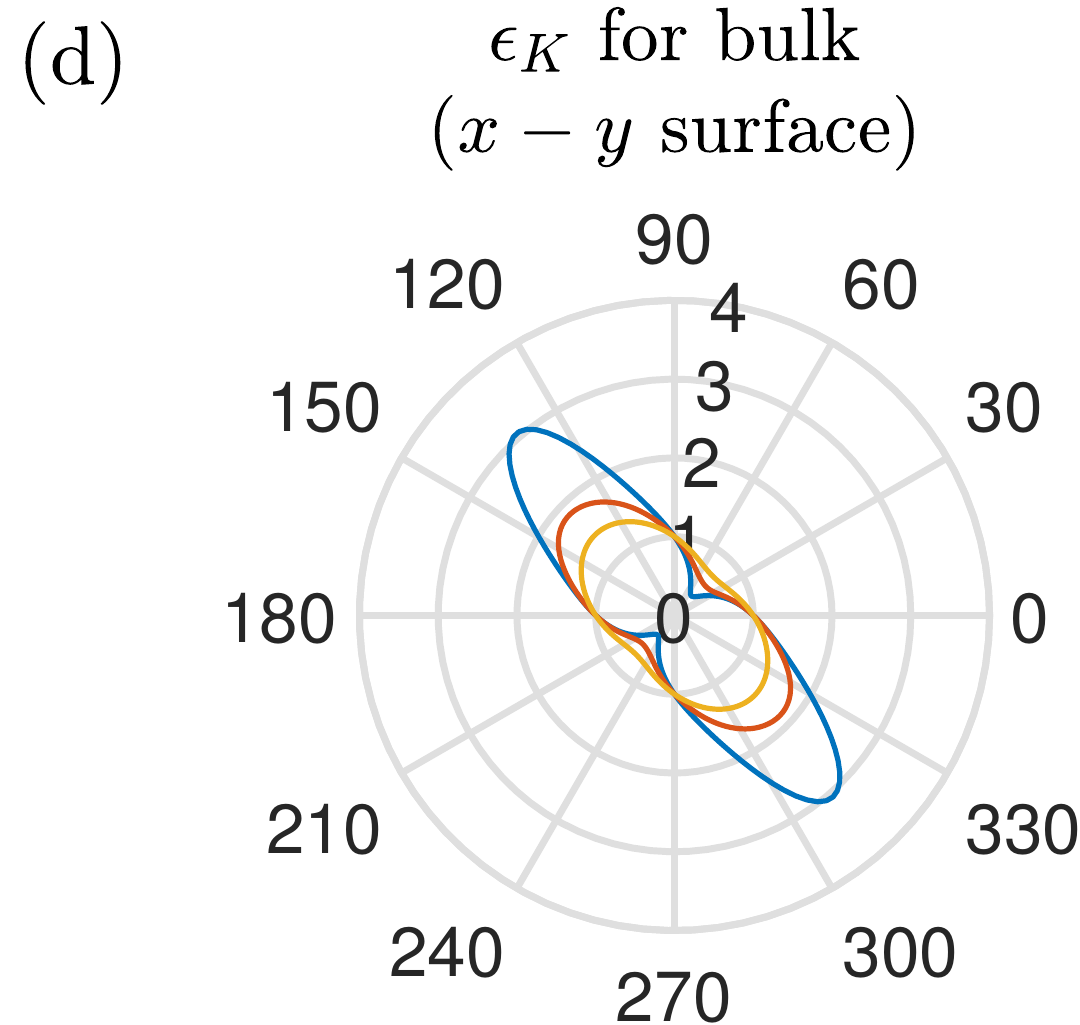}
	\includegraphics[width=0.23\textwidth]{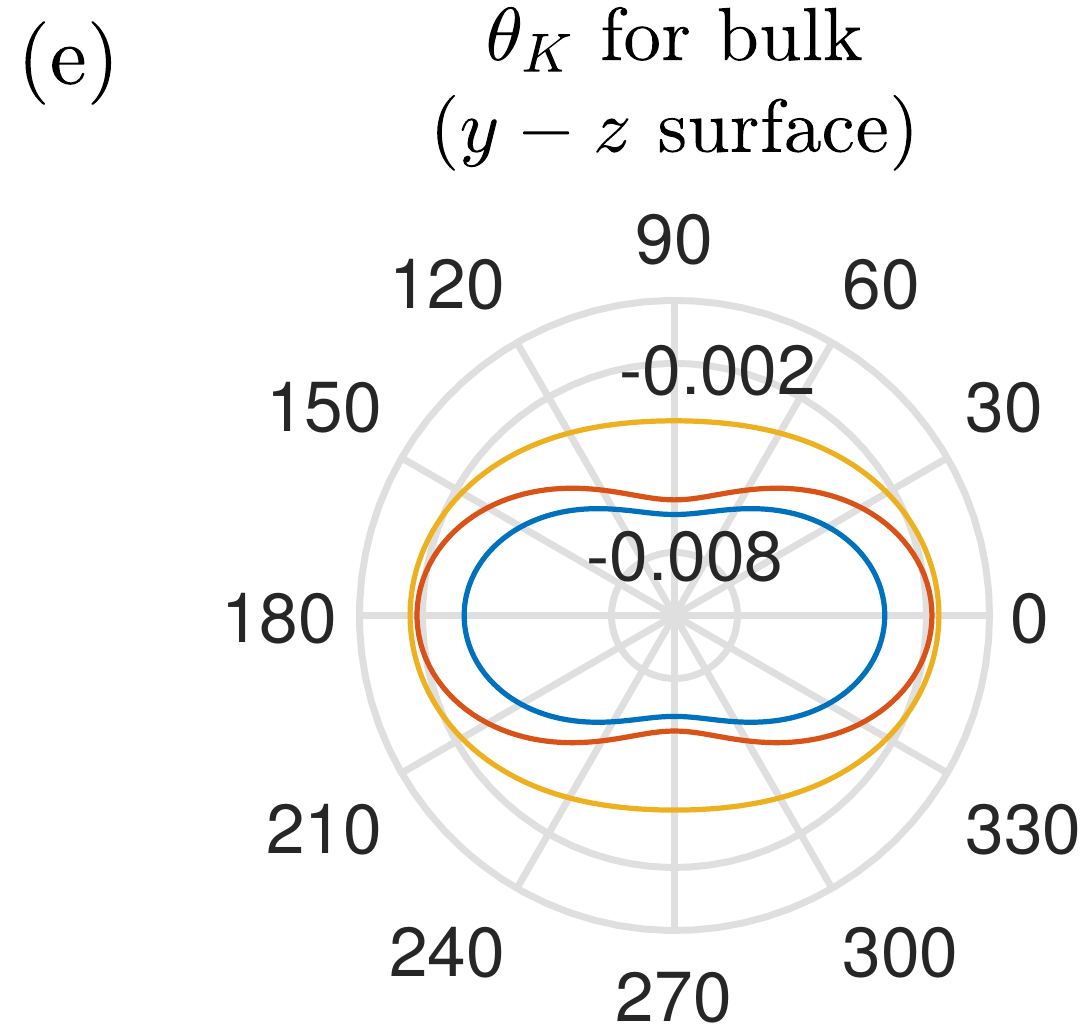}
	\includegraphics[width=0.23\textwidth]{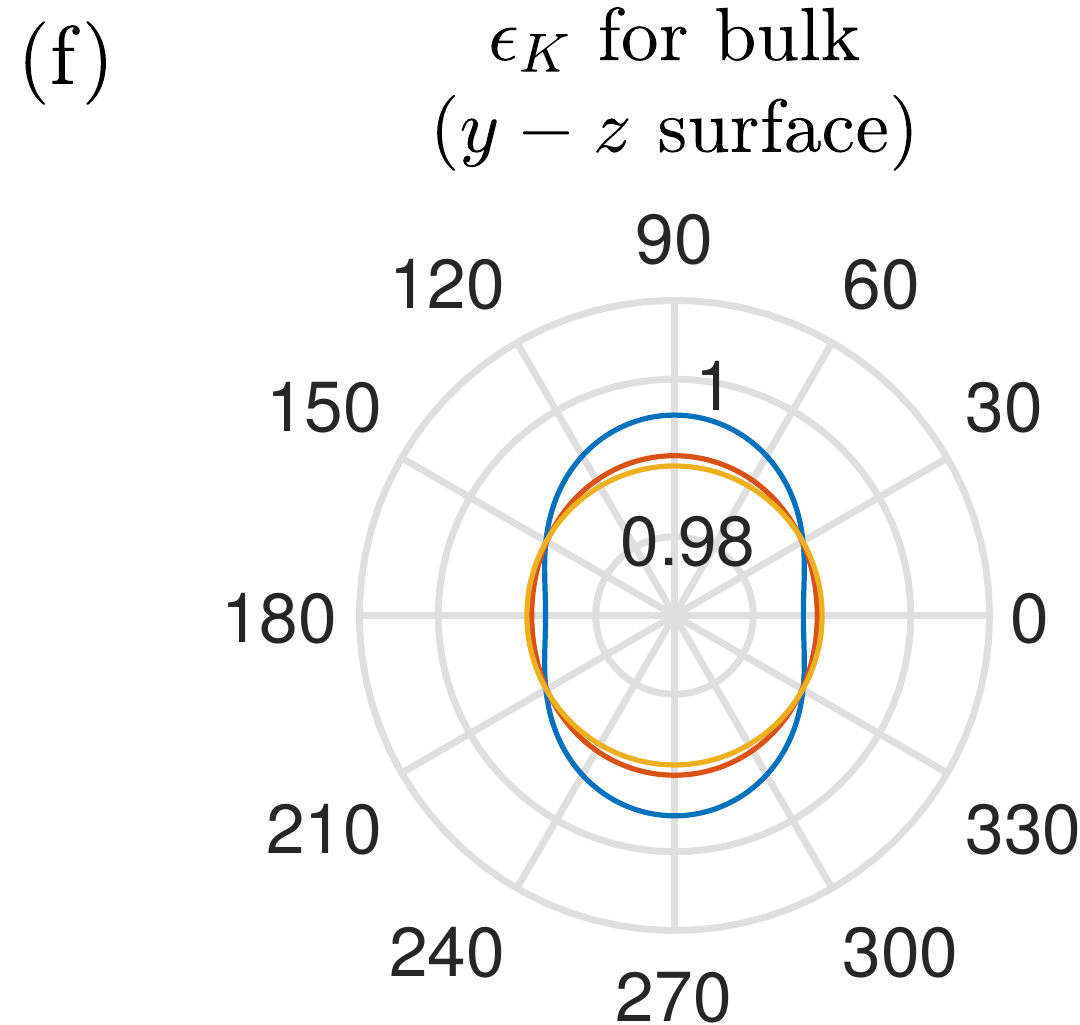}
	\caption{(a) Kerr angle and (b) Kerr ellipticity for a thin film.  (c) Kerr angle and (d) Kerr ellipticity for the bulk material with light incident on the $ y-z $ surface. (e) Kerr angle and (f) Kerr ellipticity for the bulk material and light incident on the $ y-z $ surface. The chemical potential and the tilt velocity are fixed to $\tilde{\mu} = 0$ and $ \tilde{u} = 0.1 $, respectively. For all plots the dimensionless frequency $\tilde{\omega}$ is set to the values $ \tilde{\omega} = 0.17 $ (blue), $\tilde{\omega} = 0.2$ (red), and $\tilde{\omega} = 0.23$  (yellow).}
	\label{Fig:Kerr_theta}
\end{figure}

So far we have fixed the tilt angle of the nodal loop to $\theta = \pi/4$. In this section we briefly present further numerical results about the Kerr signal when varying $ \theta $. The results are plotted in Fig.~\ref{Fig:Kerr_theta}. We first focus on the $x$-$y$ surface, Fig~\ref{Fig:Kerr_theta}(a)-(d). As seen from the figure, the Kerr rotation and ellipticity depends highly on the tilt angle. The effect is especially noticeable in the ellipticity. We observe that the ellipticity reaches large positive values in the second and fourth quadrant whereas it almost disappears in the first and third quadrant. This is because $ E_{r}^{R}(\theta) = E_{r}^{L}(\theta + \pi/2) $ as a consequence of $ \sigma_{xy}(\theta) = -\sigma_{xy}(\theta + \pi/2) $ and hence if $ \epsilon_{K} $ is large in one quadrant it has to be small in the next and vice versa.

Regarding the $ y$-$z $ surface, Fig.~\ref{Fig:Kerr_theta}(e) and (f), we note that the features are distinct from those observed for the $ x$-$y $ surface. As in the previous analysis the Kerr angle observed on the $ y$-$z $ surface is an order of magnitude smaller than that on the $ x$-$y $ surface. However, the tilt direction of the nodal loop plays a smaller role in this case, as it does not affect $ \epsilon_{xx,zz} $ as much as it affects $ \epsilon_{xy} $.

\section{Conclusion} \label{Sec:Conclusion}
In summary, we have studied the Kerr effect in a nodal loop semimetal and have described how the Kerr rotation can be related to different characteristics of the nodal loop, in particular the tilt of the nodal loop.

We have calculated the full optical conductivity tensor for a nodal-loop semimetal, which in turn has made it possible to determine the Kerr rotation. We have found that the tilt direction plays a dominating role in the determination of the conductivity. Depending on the tilt direction, the transverse conductivity oscillates between zero and a finite value. In contrast, the longitudinal conductivity always retains a finite value unless the system is Pauli blocked, and depending on the tilt angle varies between a linear and cubic behavior at small frequencies.

The Kerr rotations as a function of various system parameters, which we calculated for both a thin film and the bulk material, are the main results of this paper. We have found that the Kerr rotation is strongly dependent of the tilt velocity and the radius of the nodal loop, features which originate from the specific behavior of the transverse conductivity. Similarly to other topological materials, the obtained Kerr angle is generally large and could serve as an important tool for experimentally characterizing nodal-loop semimetals.

\begin{acknowledgments}
The authors acknowledge helpful discussions with Christoph Kastl and Alexander Holleitner at the initial stages of the project. All authors acknowledge support by the National
Research Fund, Luxembourg under grants ATTRACT A14/MS/7556175/MoMeSys, CORE C16/MS/11352881/PARTI, and C20/\-MS/\-14757511/\-OpenTop.
\end{acknowledgments}

\appendix

\begin{widetext}
\section{Real part of optical conductivity}\label{app:realpart}

As stated in the main text the real part of the optical conductivity is obtained from the following integrals, ($i,j \in \{x,y,z\}$)
\begin{equation}
	\text{Re}\left \lbrace \sigma_{ij}(\omega)\right \rbrace = \integral{k_{1}}{k_{2}}{d\xi} \Theta(1 - \tilde{\omega}/2) G_{ij}(\xi) + \integral{0}{k_{2}}{d\xi} \Theta(\tilde{\omega}/2 - 1) G_{ij}(\xi),
\end{equation}
where $G_{ij}(\xi) = g_{ij}^{+}(\xi) - g_{ij}^{-}(\xi)$ and
\begin{align}
	g_{xx}^{\pm} &= \frac{\Gamma}{2\pi^{2}\tilde{u}\tilde{\omega}^{2}}\left \lbrace \pi - \left [ 2\arccos\left ( \frac{2\tilde{\mu} \pm \tilde{\omega}}{2\tilde{u} \xi}\right ) - 2\cos 2\theta \left ( \frac{2\tilde{\mu} \pm \tilde{\omega}}{2\tilde{u} \xi}\right )\sqrt{1 - \left (\frac{2\tilde{\mu} \pm \tilde{\omega}}{2\tilde{u} \xi}\right )^{2}}\right ] \Theta\left [1 - \frac{2\tilde{\mu} \pm\tilde{\omega}}{2\tilde{u} \xi}\right ]\right \rbrace \nonumber \\
	& \hspace{5pt}\times \xi^{3}\sqrt{\left (\frac{\tilde{\omega}}{2}\right )^{2} - \left (1 - \xi^{2}\right )^{2} }, \label{Eq:fxx} \\
	g_{yy}^{\pm} &= \frac{\Gamma}{2\pi^{2}\tilde{u}\tilde{\omega}^{2}}\left \lbrace \pi - \left [ 2\arccos\left ( \frac{2\tilde{\mu} \pm \tilde{\omega}}{2\tilde{u} \xi}\right ) + 2\cos 2\theta \left ( \frac{2\tilde{\mu} \pm \tilde{\omega}}{2\tilde{u} \xi}\right )\sqrt{1 - \left (\frac{2\tilde{\mu} \pm \tilde{\omega}}{2\tilde{u} \xi}\right )^{2}}\right ] \Theta\left [1 - \frac{2\tilde{\mu} \pm\tilde{\omega}}{2\tilde{u} \xi}\right ]\right \rbrace \nonumber\\
	& \hspace{5pt}\times \xi^{3}\sqrt{\left (\frac{\tilde{\omega}}{2}\right )^{2} - \left (1 - \xi^{2}\right )^{2} },
	\label{Eq:fyy} \\
	g_{xy}^{\pm} &= \Gamma\frac{2\tilde{\mu} \pm\tilde{\omega}}{\pi^{2}\tilde{u}\tilde{\omega}^{2}}\sin 2\theta \xi\sqrt{\left (\frac{\tilde{\omega}}{2}\right )^{2} - \left (1 - \xi^{2}\right )^{2}}\sqrt{\xi - \left (\frac{2\tilde{\mu} \pm\tilde{\omega}}{2\tilde{u}}\right )^{2}}\Theta \left  [1 - \frac{2\tilde{\mu} \pm\tilde{\omega}}{2\tilde{u} \xi} \right ]
	\label{Eq:fxy} \\
	g_{zz}^{\pm} &= \frac{1}{\Gamma}\frac{1}{2\pi^{2}\tilde{\omega}^{2}}\left\lbrace2\pi - 4\arccos\left (\frac{2\tilde{\mu} \pm \tilde{\omega}}{2\tilde{u} \xi} \right ) \Theta \left  [ 1 -  \frac{2\tilde{\mu} \pm \tilde{\omega}}{2\tilde{u} \xi}\right ] \frac{\xi\left ( 1 - \xi^{2}\right )^{2}}{\sqrt{\left (\frac{\tilde{\omega}}{2} \right )^{2} - \left ( 1 - \xi^{2} \right )^{2} }} \right \rbrace.
	\label{Eq:fzz}
\end{align}
In the limit $\tilde{\omega} \ll 1$, $\tilde{\omega} < \tilde{u}$ and $\tilde{\mu} = 0$ we can obtain approximate results,
\begin{align}
\text{Re}\lbrace \sigma_{xx}(\omega) \rbrace &\approx \Gamma\frac{1}{2\pi^{2}\tilde{u}}\left[\frac{5}{12}\left (1 - \cos 2\theta\right )\tilde{\omega} + \frac{40}{2304\tilde{u}^{2}}\left (1 + 3\cos 2\theta\right )\tilde{\omega}^{3} - \frac{39}{2304\tilde{u}}\left (1 - \cos 2\theta\right )\tilde{\omega}^{3}\right], \\
\text{Re}\lbrace \sigma_{yy}(\omega) \rbrace  &\approx \Gamma\frac{1}{2\pi^{2}\tilde{u}}\left[\frac{5}{12}\left (1 + \cos2\theta \right )\tilde{\omega} + \frac{40}{2304\tilde{u}^{2}}\left (1 - 3\cos 2\theta\right )\tilde{\omega}^{3} - \frac{39}{2304\tilde{u}}\left (1 + \cos 2\theta\right )\tilde{\omega}^{3}\right], \\
\text{Re}\lbrace \sigma_{xy}(\omega) \rbrace &\approx \Gamma\frac{2\sin 2\theta}{\pi^{2} \tilde{u}} \left[\frac{5}{24}\tilde{\omega} - \frac{40 + 13\tilde{u}}{1536\tilde{u}}\tilde{\omega}^{3}\right],  \\
\text{Re}\lbrace \sigma_{zz}  (\omega) \rbrace &\approx \frac{1}{12\pi^{2}\tilde{u}\Gamma} \left[  5\tilde{\omega}  + \frac{26 + 27\tilde{u}^{2}}{120\pi^{2}\tilde{u}^{2}\Gamma}\tilde{\omega}^{3}\right].
\end{align}
The linear terms are reported in Eqs.~\eqref{Eq:app_sxx} to \eqref{Eq:app_szz}.

\section{Intraband transitions} \label{App:Intraband}
Here we give the functions $ \tilde{G}_{ij} $ that enters the calculation of the intraband amplitude $ D_{ij}(\omega) $ in Eq.~\eqref{Eq:IntrabandAmplitude},
\begin{align}
	\tilde{G}_{xx} &= \omega_{0}\Gamma\xi\frac{\tilde{\mu} - \tilde{u}\xi\cos(\phi - \theta) }{\sqrt{(\tilde{\mu} - \tilde{u}\xi\cos(\phi - \theta))^{2}  - (1 - \xi^{2})^{2}}}\left[\tilde{u}\cos\theta + \frac{2\xi \cos\phi (1-\xi^{2})}{\tilde{\mu} - \tilde{u}\xi\cos(\phi - \theta) }\right]^{2} \Theta\left[ (\tilde{\mu} - \tilde{u}\xi\cos(\phi - \theta))^{2}  - (1 - \xi^{2})^{2} \right], \\
	\tilde{G}_{yy} &= \omega_{0}\Gamma  \xi\frac{\tilde{\mu} - \tilde{u}\xi\cos(\phi - \theta) }{\sqrt{(\tilde{\mu} - \tilde{u}\xi\cos(\phi - \theta))^{2}  - (1 - \xi^{2})^{2}}}\left[\tilde{u}\sin\theta + \frac{2\xi \sin\phi (1-\xi^{2})}{\tilde{\mu} - \tilde{u}\xi\cos(\phi - \theta) }\right]^{2} \times \Theta\left[ (\tilde{\mu} - \tilde{u}\xi\cos(\phi - \theta))^{2}  - (1 - \xi^{2})^{2} \right],\\
	\tilde{G}_{xy} &= \omega_{0}\Gamma \xi\frac{\tilde{\mu} - \tilde{u}\xi\cos(\phi - \theta) }{\sqrt{(\tilde{\mu} - \tilde{u}\xi\cos(\phi - \theta))^{2}  - (1 - \xi^{2})^{2}}} \nonumber \\
	&\hspace{50pt} \times \left[\tilde{u}\sin\theta + \frac{2\xi \sin\phi (1-\xi^{2})}{\tilde{\mu} - \tilde{u}\xi\cos(\phi - \theta) }\right] \left[\tilde{u}\cos\theta + \frac{2\xi \cos\phi (1-\xi^{2})}{\tilde{\mu} - \tilde{u}\xi\cos(\phi - \theta) }\right]  \Theta\left[ (\tilde{\mu} - \tilde{u}\xi\cos(\phi - \theta))^{2}  - (1 - \xi^{2})^{2} \right], \\
	\tilde{G}_{zz} & = \frac{\omega_{0}}{\Gamma}\xi\frac{\sqrt{(\tilde{\mu} - \tilde{u}\xi\cos(\phi - \theta))^{2}  - (1 - \xi^{2})^{2}}}{\sqrt{(\tilde{\mu} - \tilde{u}\xi\cos(\phi - \theta))^{2}}} \Theta\left[ (\tilde{\mu} - \tilde{u}\xi\cos(\phi - \theta))^{2}  - (1 - \xi^{2})^{2} \right]	
\end{align}

\end{widetext}

\bibliography{bibliography}

\end{document}